\documentclass[12pt,letterpaper]{article}
\usepackage{graphicx}
\usepackage{amsmath,amsthm,amssymb,amsfonts}

\usepackage{tikz}

\usepackage[T1]{fontenc}
\usepackage{calligra}

\newcommand{\bC}{{\bf C}}

\newcommand{\bH}{{\bf H}}
\newcommand{\bI}{{\bf I}}
\newcommand{\bM}{{\bf M}}

\newcommand{\bR}{{\bf R}}
\newcommand{\bS}{{\bf S}}
\newcommand{\bU}{{\bf U}}

\newcommand{\bX}{{\bf X}}

\newcommand{\cF}{{\cal F}}

\newcommand{\cH}{{\cal H}}

\newcommand{\cK}{{\cal K}}

\newcommand{\cM}{{\cal M}}

\newcommand{\cT}{{\cal T}}

\newcommand{\mC}{\mathbb{C}}

\newcommand{\mP}{\mathbb{P}}
\newcommand{\mQ}{\mathbb{Q}}

\newtheorem{thm}{Theorem}

\theoremstyle{remark}

\begin{document}
\title{Learning Relationship between Quantum Walks and Underdamped Langevin Dynamics}  
\author{Yazhen Wang  
\\ University of Wisconsin-Madison}
\maketitle

\begin{abstract}

Fast computational algorithms are in constant demand, and their development has been driven by advances such as quantum speedup and classical acceleration. This paper intends to study 
search algorithms based on quantum walks in quantum computation and sampling algorithms based on Langevin dynamics in classical computation. On the quantum side, quantum walk–based search algorithms can achieve quadratic speedups over their classical counterparts. In classical computation, a substantial body of work has focused on gradient acceleration, with gradient-adjusted algorithms derived from underdamped Langevin dynamics providing quadratic acceleration over conventional Langevin algorithms.

Since both search and sampling algorithms are designed to address learning tasks, we study learning relationship  
between coined quantum walks and underdamped Langevin dynamics. 
Specifically, we show that, in terms of the Le Cam deficiency distance, a quantum walk with randomization is asymptotically equivalent to underdamped Langevin dynamics, whereas the quantum walk without randomization is not asymptotically equivalent due to its high-frequency oscillatory behavior. We further discuss the implications of these equivalence and nonequivalence results for the computational and inferential properties of the associated algorithms in machine learning tasks. Our findings offer new insight into the relationship between quantum walks and underdamped Langevin dynamics, as well as the intrinsic mechanisms underlying quantum speedup and classical gradient acceleration. 

\end{abstract}

\section{Introduction} 
\label{SEC-Intro}

The current computing power demand in machine learning pushes the limits of computer technology, and great efforts are devoted to develop fast computational techniques from chips to software to systems. Notably, there has been surging interest in and great research work on developing faster algorithms in both classical computation and quantum computation. Here we focus on two independent strands of such research: a classical approach and a quantum approach. The classical approach is accelerated gradient-based optimization methods, and the quantum approach is quantum walk-based search algorithms. Both approaches can provide quadratically faster algorithms than their corresponding standard counterparts. See  \cite{grover1996fast} \cite{grover1997quantum} \cite{nielsen2000quantum} 
\cite{portugal2018} \cite{wang2022harvardquantum}
for quantum computation and \cite{ma2021there} \cite{nesterov2013introductory} \cite{nesterov1983method} \cite{su2016} \cite{wang2020asymptotic} \cite{wibisono2016variational} for classical computation. 
This paper studies the relationship of these previously unrelated classical and quantum algorithms from a learning perspective and establishes learning equivalence and nonequivalence between the algorithms when applying to inferential tasks in machine learning. 

In quantum computation, the quantum walk search is a quantum algorithm designed to search via quantum walk for a marked node within a graph. The concept of a quantum walk is derived from classical random walks, where a walker moves randomly through a graph or lattice. In a classical random walk, the walker's position is described by a probability distribution over the graph's nodes. In contrast, a quantum walk represents the walker as a quantum state, which can be in a superposition of multiple locations simultaneously. The computational efficiency and success probability of these algorithms are strongly influenced by the structure of the search space. Typically, quantum walk search algorithms provide a quadratic speedup compared to their classical counterparts. 
For instance, Grover's algorithm---a well-known quantum search algorithm for unstructured search---can find a solution with 
complexity of order $O(\sqrt{N})$, where $N$ represents the size of the search problem. In classical computation, solving the same problem would require a complexity of order $O(N)$. Moreover, the computational complexity results of orders $O(N)$ and $O(\sqrt{N})$ represent the theoretical limits for solving the search problem. Specifically, any quantum algorithm must perform at least $O(\sqrt{N})$ operations, while classical algorithms cannot solve the problem with fewer than $O(N)$ operations. Consequently, Grover's algorithm is asymptotically optimal, providing a quadratic speedup 
over the best classical search algorithm. See \cite{grover1996fast} \cite{grover1997quantum}  \cite{magniez2007} \cite{portugal2018} \cite{santha2008}
\cite{shenvi2003} for details.

In the context of classical computation, 
Nesterov's accelerated gradient descent method, introduced by Nesterov in 1983 (see \cite{nesterov1983method}), 
is a first-order algorithm designed to minimize convex objective functions. It designs an ingenious algorithmic scheme to incorporate an additional variable called momentum and accelerate conventional gradient descent. Specifically, for convex functions, the convergence rate of Nesterov's method is $O(k^{-2})$, where $k$ is the number of iterations, compared to $O(k^{-1})$ 
for conventional gradient descent. The concept of acceleration has become central to gradient-based optimization methods. By allowing the step size to approach zero, Nesterov's accelerated gradient method leads to a second-order ordinary differential equation (ODE), referred to as Nesterov's accelerated gradient flow or Nesterov's ODE. See \cite{nesterov2013introductory} \cite{nesterov1983method} \cite{su2016}  \cite{wang2020asymptotic} \cite{wibisono2016variational} for details.

A seminal work by Jordan, Kinderlehrer, and Otto in 1998 (see \cite{jordan1998variational})  
established that the Fokker-Planck equation 
of Langevin dynamics corresponds to the gradient flow of the relative entropy functional in the space of probability measures equipped with the Wasserstein-$2$ metric, which refers to as Wasserstein gradient flow. This insight has forged a link between the fields of sampling and optimization by studying optimization problems in the Wasserstein-$2$ space. Recent studies have extended the Wasserstein gradient flow to 
Nesterov's accelerated gradient flow, exploring the acceleration in sampling from Langevin dynamics. 
An accelerated version of Langevin dynamics, known as underdamped Langevin dynamics, shares key similarities with Nesterov's accelerated gradient flow. Underdamped Langevin dynamics combines a deterministic component, which directly corresponds to Nesterov's accelerated gradient flow, and a stochastic component. This approach enables accelerated Langevin dynamics to achieve a sampling speed that is quadratically faster than conventional Langevin dynamics.
See 
\cite{ambrosio2008gradient}
\cite{cheng2018a}
\cite{cheng2020}
\cite{jordan1998variational} 
\cite{ma2021there}
\cite{pavliotis2014} 
\cite{zuo2024} for details. 

Computational and inferential tasks are two key components in machine learning. Computationally, underdamped Langevin dynamics and quantum walk both can 
provide a quadratic speedup over their corresponding standard counterparts. In this paper we use Le Cam's deficiency distance to investigate their relationship from the inferential learning perspective. Specifically, we put learning algorithms (or methods)  based on quantum walk and underdamped Langevin dynamics in the Le Cam paradigm and establish their asymptotic equivalence and nonequivalence in terms of  Le Cam's deficiency distance. Here equivalence means that of the two learning models associated with the quantum walk and 
underdamped Langevin dynamics, each 
 learning procedure for one model has a corresponding equal-performance 
 learning procedure for another model, and asymptotic equivalence signifies the equal-performance in an asymptotic sense. See \cite{lecam1986} \cite{wang2002} \cite{wang2013asymptotic} for details. 

Our study of learning relationships may shed new light on acceleration phenomena in classical computation and quantum speedup in quantum computation, and it motivates us to explore whether 
there exist intrinsic connections between classical acceleration and quantum speedup. A one-dimensional quantum walk has a corresponding one-dimensional classical random walk; however, the quantum walk exhibits a mixing time that is quadratically faster than that of its classical counterpart. We can lift the classical random walk to a two-dimensional classical random walk that 
matches the quantum walk in terms of mixing time.

Several parallels emerge from this comparison. First, both quantum walks and underdamped Langevin dynamics exhibit quadratic acceleration in mixing time relative to their respective classical counterparts---namely, classical random walks and standard (overdamped) Langevin dynamics. Specifically, a quantum walk may achieve a mixing time that is quadratically faster than that of the corresponding classical random walk, while algorithms based on underdamped Langevin dynamics can enjoy a quadratically faster mixing time than those based on standard Langevin dynamics.

Second, lifting mechanisms underlie the quadratic acceleration observed in both settings. On the quantum side, a quantum walk can be interpreted as corresponding to a classically lifted Markov chain, which accelerates mixing and thus reduces computational complexity. On the classical side, since Nesterov’s gradient flow is governed by a second-order ordinary differential equation (ODE), the underdamped Langevin dynamics can be viewed as a second-order stochastic differential equation (SDE). It lifts the standard Langevin dynamics from a one-dimensional space to a two-dimensional process consisting of the original position variable and an additional momentum variable, resulting in a two-dimensional SDE. 

We hope that this line of research will stimulate further investigation into the connections between quantum speedup and classical acceleration, as well as deeper exploration of their shared mechanisms and underlying principles.

The rest of paper proceeds as follows. Section 2 begins with a brief review of quantum walks and Langevin dynamics, and then presents the main results on 
the asymptotic equivalence and nonequivalence of quantum walks and underdamped Langevin dynamics, along with a discussion of their implications. 
Section 3 features concluding remarks, while the appendix collects all technical proofs. 

\section{Main results on quantum walk and Langevin dynamics} 
\label{SEC-revew} 

\subsection{Quantum walk}
\label{SEC-QW} 

Since this paper focuses on coined discrete-time quantum walks on the real line, we provide a brief review of the framework. Such a quantum walk consists of a walker, a coin, evolution operators for both the walker and the coin, and a set of observables. 

The walker is a quantum system that resides in a Hilbert space $\cH_p$ of an infinite but countable dimension. It is customary to represent the walker's position 
using the canonical computational basis of $\cH_p$, whose basis vectors correspond to its discrete position sites. We denote the walker's state by 
$| \mbox{position} \rangle \in \cH_p$ and note that 
the canonical basis states $\{| i \rangle_p\}$ span $\cH_p$. Any superposition of the form $\sum_i \alpha_i | i \rangle_p$, subject to $\sum_i |\alpha_i|^2=1$,
is therefore a valid state of the walker. 
The walker is typically initialized at the origin, that is, $| \mbox{position} \rangle_{\mbox{initial}} = | 0 \rangle_p$. 

The coin is a quantum system living in a two-dimensional Hilbert space $\cH_c$. It may take either of the canonical basis states $| 0 \rangle_c$ and $| 1 \rangle_c$, 
as well as any superposition of these basis states. Thus, the coin state $| \mbox{coin} \rangle \in \cH_c$, and a general normalized coin state can be written as
$| \mbox{coin} \rangle = a_0 | 0 \rangle_c + a_1 | 1 \rangle_c$, where $| a_0 |^2 + |a_1|^2 = 1$.

The total state of the quantum walk resides in $\cH = \cH_p \otimes \cH_c$. It is customary to choose product states in $\cH$ as initial states; that is, 
$| \psi \rangle_{\mbox{initial}}= | \mbox{position} \rangle_{\mbox{initial}} \otimes |\mbox{coin}\rangle_{\mbox{initial}}$. 

The evolution of the quantum walk is governed by two operators acting on the coin and the walker. At each step, a coin operator is first applied to the coin state, followed by a conditional shift operator acting on the combined system. The role of the coin operator is to transform the coin state, while randomness is introduced through measurements performing on the total system after repeated applications of both evolution operators. 

Among coin operators, customarily denoted by $\bC$, the Hadamard operator $\bH$ has been used extensively:
\begin{equation}\label{Hadamard-operator}
    \bH = \frac{1}{\sqrt{2}} \left( | 0 \rangle_c \langle 0 | + | 0 \rangle_c  \langle 1 | + | 1\rangle_c \langle 0| - | 1 \rangle_c \langle 1 | \right). 
\end{equation} 
For the conditional shift operator, a unitary operator is employed that moves the walker one step forward when the associated coin state is one of the 
basis states (e.g. $| 0 \rangle$), and one step backward when the coin state is the 
other basis state (e.g. $ | 1 \rangle$). A suitable conditional shift operator is given by 
\begin{equation} \label{shift-operator} 
  \bS = | 0 \rangle_c \langle 0 | \otimes \sum_j | j+1 \rangle_p \langle j | + | 1 \rangle_c \langle 1 | \otimes \sum_j | j-1 \rangle_p \langle j | .
\end{equation} 
Consequently, the evolution operator acting on the total Hilbert space is given by $\bU = \bS \cdot (\bC \otimes \bI_p)$, where $\bI_p$ denotes the identity operator on $\cH_p$. 
A concise mathematical representation of the quantum walk's state after $t$ steps is then given by 
\[ | \psi \rangle_t = \bU^t | \psi \rangle_{\mbox{initial}}, \]
where $| \psi \rangle_{\mbox{initial}} = | \mbox{position} \rangle_{\mbox{initial}} \otimes |\mbox{coin} \rangle_{\mbox{initial}}$.
A quantum walker is referred to as a Hadamard walk if the two evolution operators employed are the Hadamard operator $\bH$ and the conditional shift operator $\bS$ defined above. 

Observables are needed to perform measurements and obtain the outcomes of the  quantum walk. We define a set of
observables based on the basis states used to describe the coin and the walker. There are several ways to extract information from the composite quantum system. For example, 
one may first perform a measurement on the coin using the observable
\[ \bM_c = a^w_{0} | 0 \rangle_c \langle 0 | + a^w_1 | 1 \rangle_c \langle 1| . \] 
Subsequently, a measurement on the walker's position state is carried out using the operator
\[ \bM_p = \sum_i \alpha^w_{i} | i \rangle_p \langle i | . \]

We may express the state $ | \psi \rangle_t$ in the form of 
\[ | \psi \rangle_t = \sum_k \left[ b_{kt} |0 \rangle_c + \beta_{kt}  |1 \rangle_c  \right] | k \rangle_p , \] 
where $| 0 \rangle_c$ and $|1 \rangle_c$ denote the coin-state components and $| k \rangle_p$ denote the position-state
components. For example, suppose the quantum walk is initialized in the state  
\[ | \psi \rangle_0 = | 0 \rangle_c \otimes |0 \rangle_p , \]
with equations (\ref{Hadamard-operator}) and (\ref{shift-operator}) defining the coin and shift operators, respectively. In this case, one can derive 
analytic expressions for $b_{kt}$ and $\beta_{kt}$, as well as for the probability that the walker is found at position $k$ when a measurement is performed at the $t$-th step.

Let $\mathbb{C}$ denote the set of all complex numbers, and let $\varphi = (a_0, a_1)^\prime \in \mathbb{C}^2$, with $ |a_0|^2+|a_1|^2=1$, represent an initial coin state of a one-dimensional coined quantum walk. Let 
$\bX^q_t$ denote the one-dimensional coined quantum walk at time $t$, starting from an initial coin state  $\varphi$ and  evolving under a coin operator $\bC$ described by a $2 \times 2$ 
unitary matrix $\bU$. 

This quantum walk 
exhibits a linear standard deviation of order $O(t)$ and a mixing time of order $O(t)$, in sharp contrast to the corresponding classical random walk, which has a standard deviation of order $O(\sqrt{t})$ and a mixing time of order $O(t^2)$. See \cite{ambainis2004} \cite{apers2018} \cite{kadian2021} \cite{kempe2003} \cite{magniez2007}  \cite{portugal2018} \cite{santha2008} \cite{shenvi2003} \cite{venegas2012} \cite{wong2017} for more details.

\subsection{Langevin dynamics} 
\label{SEC-Langevin} 
The underdamped Langevin dynamics is defined by the following SDE,
\begin{equation}  \label{underdamped}
  \left\{ \begin{array}{l}
      d Y^c_t = V^c_t dt ,  \qquad (Y^c_0, V^c_0) \sim \rho_0, \\ 
      d V^c_t = - V^c_t dt - \nabla \Psi(Y^c_t) dt + \sqrt{ 2 }  dB_t  , 
      \end{array} 
\right. 
\end{equation} 
where $B_t$ denotes a standard one-dimensional Brownian motion,  and $\Psi(\cdot)$ is a potential function, $\nabla$ denotes the gradient operator, and $\rho_0$ is an initial probability density function (pdf).  
The solution $(Y^c_t, V^c_t)$ to this SDE is a diffusion process with a unique invariant distribution proportional to $\exp( - \Psi(x) - v^2/2 )$. Consequently,  the marginal invariant distribution of $Y^c_t$ is proportional to $\exp( - \Psi(x) )$.  An algorithm based on a discretization of the underdamped Langevin equation (\ref{underdamped}) yields a quadratically faster method 
for sampling from the invariant distribution compared to conventional Langevin algorithms derived from the standard (or overdamped) Langevin dynamics.  
See \cite{cao2023}
\cite{cheng2018a}
\cite{eberle2019}
\cite{pavliotis2014} 
\cite{zuo2024} for more details.

\subsection{Learning equivalence and nonequivalence} 
\label{SEC-equiv} 

A classical random walk can be obtained by discretizing the standard Langevin dynamics, which itself may be viewed as a continuous analogue of the classical random walk. Indeed, the classical random walk and the standard Langevin dynamics are asymptotically equivalent in terms of mathematical approximation and statistical learning. 
Both quantum walks and underdamped Langevin dynamics exhibit quadratically faster mixing times than their respective counterparts---the classical random walk and standard Langevin dynamics---and the associated algorithms achieve quadratic speedups over those based on classical random walks and standard Langevin dynamics.

Moreover, quantum walks and underdamped Langevin dynamics can each be related to their classical counterparts through appropriate lifting schemes. Specifically, a quantum walk corresponds to a lifted classical random walk, while underdamped Langevin dynamics can be viewed as a lifted version of standard Langevin dynamics. This naturally raises the question of whether there exists a deeper relationship between quantum walks and underdamped Langevin dynamics from a learning or algorithmic perspective.

As described in Section \ref{SEC-QW},  $\bX^q_t$ denotes the coined quantum walk at time $t$ with an initial coin state $\varphi$ 
and an evolution operator given by unitary matrix $\bU$, as described in Section \ref{SEC-QW}, where $\theta=(\varphi, \bU)$, 
\[ \varphi =  \begin{pmatrix} a_0 \\ a_1 \end{pmatrix} \in \mathbb{C}^2 \mbox{ and }  \bU = \begin{pmatrix} u_{11} & u_{12} \\ u_{21} & u_{22} \end{pmatrix} \in \mC^{2\times 2} 
\]
satisfy $ |a_0|^2+|a_1|^2=1$, $\bU^\dagger \bU= \bI_2$, and $u_{11} u_{12} u_{21} u_{22} \neq 0$. 
Here $*$ and $\dagger$ denote conjugate and conjugate transpose, respectively, $|a|^2 = a^* a$, and $\bI_2$ denotes the two-dimensional identity matrix.
Let $X^q_t$ denote the position measurement outcome of $\bX^q_t$. For an arbitrarily small $\eta>0$, we define $X^q_t(\eta)$ to be a random variable taking a value $X^q_t+j$, 
$j=0, \pm 1, \cdots, \pm (r-1), r$, with probability $1/ (2 r)$, where $r=[\eta t]$. 
Denote by $\mP^q_{t,\theta}$ and $\mP^q_{t,\theta,\eta}$ the probability distributions of $X^q_t$ and $X^q_t(\eta)$, respectively. 
We refer to $X^q_t(\eta)$ as a randomization of $X^q_t$. Randomization is a statistical technique widely applied in various fields, especially in research studies and scientific applications, 
to enhance the validity of statistics and machine learning methods.

Denote by $\mP^c_{t,\theta, \epsilon}$ the probability distribution of $X^c_t(\epsilon) = t |u_{11} | e^{- | Y ^c_t|} sign(Y^c_t)$, where the classical process $Y^c_t$ is 
governed by the underdamped Langevin dynamics (\ref{underdamped}) with the potential function $\Psi_\epsilon (y) = - \log \pi_\epsilon(y; \theta)$, where $\epsilon$ is an arbitrarily small positive constant, $\pi_\epsilon(y; \theta) = \pi(y; \theta)$ for $|y| \geq -\log(1-\epsilon)$ and $\pi_\epsilon(y; \theta) = \varkappa_\epsilon$ for $|y| < -\log(1-\epsilon)$, 
\begin{eqnarray}\label{varkappa}
  \varkappa_\epsilon &=& \frac{1}{-2\log(1- \epsilon)} \left[ 1 -  \int_{|y| \geq -\log(1-\epsilon)} \pi_\epsilon(y; \theta) dy \right]  \nonumber \\
  &=& \frac{1}{-2\log(1- \epsilon)} \int_{\log(1-\epsilon)}^{-\log(1-\epsilon)} \pi(y; \theta) dy ,
 \end{eqnarray}
and $\pi(y; \theta)$ is a pdf defined for $y \neq 0$ as follows, 
\begin{eqnarray} \label{Konno1}
&& \pi(y; \theta) = \frac{ e^{-|y|}\sqrt{ 1- |u_{11}|^2} } {\pi (1 - |u_{11}|^2 e^{-2 |y|} ) \sqrt{ 1 - e^{-2 |y|} } }  \nonumber \\
&&  \left\{ 1 - \left( |a_0|^2  - |a_1|^2 + \frac{ a_0 a_1^* u_{11} u_{12}^* + a_0^* a_1 u_{11}^* u_{12} 
} {|u_{11}|^2} \right)  |u_{11}| e^{-|y|}  \mbox{sign}(y) \right\} . \qquad 
\end{eqnarray}
Here sign$(y)$ denotes the sign function. 

Let $(\Omega_q, \cF_q)$ and $(\Omega_c, \cF_c)$ be the measurable spaces where $\mP^q_{t, \theta,\eta}$ and $\mP^c_{t,\theta,\epsilon}$ live, respectively. We define a Markov kernel $\cK(\cdot, \cdot)$ to map $\mP^c_{t,\theta, \epsilon}$ on $(\Omega_c, \cF_c)$ into another probability distribution on 
$(\Omega_q, \cF_q)$ so that the mapped $\mP^c_{t, \theta, \epsilon}$ can be compared with $\mP^q_{t,\theta,\eta}$. Specifically, for a given $\omega \in \Omega_c$, $\cK(\omega, \cdot)$ is a probability measure on the $\sigma$-field $\cF_q$, and for a fixed $A \in \cF_q$, $\cK(\cdot, A)$ is a measurable function on $\Omega_c$. The Markov kernel
$\cK(\omega, A)$ is defined for $\omega \in \Omega_c$ and $A \in \cF_q$ and maps $\mP^c_{t,\theta,\epsilon}$ into a probability measure, denoted by $\cK(\mP^c_{t,\theta, \epsilon})$, on $(\Omega_q, \cF_q)$ as follows: For any $A \in \cF_q$, we define 
\[   [\cK(\mP^c_{t,\theta,\epsilon})](A) = \int_{\Omega_c}  \cK(\omega, A) \mP^c_{t,\theta,\epsilon}(d\omega) . \]
The deficiency of $X ^c_t(\epsilon) \sim \mP^c_{t,\theta,\epsilon}$ with respect to $X^q_t(\eta) \sim \mP^q_{t,\theta,\eta}$ is defined by 
\[ \gimel (X^c_{t}(\epsilon), X^q_{t}(\eta)) = \inf_{\cK} \sup_{\theta \in \Theta} \| \mP^q_{t,\theta,\eta} - \cK(\mP^c_{t,\theta,\epsilon}) \|_{TV} , \]
where $\| \cdot \|_{TV}$ is the total variation norm, the infimum is over all Markov kernels, the supremum is over 
some parameter space $\Theta$. 
 Similarly, the deficiency $\gimel(X^q_{t}(\eta), X^c_{t}(\epsilon))$ of $X^q_{t}(\eta)$ with respect to $X^c_{t}(\epsilon)$ can be defined. We define Le Cam deficiency distance between $X^q_t(\eta)$ and $X^c_t(\epsilon)$ as 
\[  \mho(X^q_t(\eta), X^c_t(\epsilon)) 
     = \max \{ \gimel(X^q_{t}(\eta), X^c_{t}(\epsilon)),  \gimel (X^c_{t}(\epsilon), X^q_{t}(\eta))  \} .  \]  
 Also, using the distribution $\mP^q_{t,\theta}$ of $X^q_t$, we can define $\gimel (X^c_{t}(\epsilon), X^q_{t})$, $ \gimel (X^q_{t}, X^c_{t}(\epsilon))$, and $\mho(X^q_t, X^c_t(\epsilon))$ (see \cite{lecam1986} \cite{wang2002} \cite{wang2013asymptotic}).  
  
We need to fix some notations and define the parameter space $\Theta$ for the theoretical analysis. We rewrite $\varphi$ and $\bU$ as
\begin{equation} \label{equ-U}
 \varphi =  \begin{pmatrix} e^{i\varsigma_1} \cos\vartheta \\ e^{i\varsigma_2} \sin \vartheta \end{pmatrix} ,  
 \bU = \begin{pmatrix} e^{i \gamma_1} \cos \phi  & e^{i\gamma_2} \sin \phi  \\ e^{-i \gamma_2} \sin \phi  & - e^{-i\gamma_1} \cos \phi \end{pmatrix}, 
 \end{equation} 
 where $\varsigma_1, \varsigma_2 , \gamma_1, \gamma_2 \in [-\pi/2, \pi/2]$, and $\vartheta, \phi \in [0, \pi/2]$. 
Define 
\begin{equation} \label{Theta} 
 \Theta = \{ \theta=(\varphi, \bU):  \varphi \mbox{ and } \bU \mbox{ are given in (\ref{equ-U}) with } 
  \phi \in [\kappa,\pi/2-\kappa] \}   , 
\end{equation} 
where $\kappa$ is a small positive constant. 

We have the following theorem on the Le Cam distance between the quantum walk and underdamped Langevin dynamics. 
\begin{thm} Under the condition (\ref{Theta}), we have 
\begin{equation} \label{mho-1}
   \limsup_{t \rightarrow \infty} \mho ( X^q_{t}(\eta), X^c_t(\epsilon))  = O(\epsilon^{1/2} + \eta^{1/2} )    , 
 \end{equation} 
 which goes to zero as $\epsilon$ and $\eta \rightarrow 0$.
Furthermore, we can show 
\begin{equation} \label{n-X-qc} 
\liminf_{t \rightarrow \infty} \mho (X^q_t, X^c_t(\epsilon) )  
\geq c_0 - O(\epsilon^{1/2})  ,
\end{equation} 
which is bounded below by $c_0$ as $\epsilon \rightarrow 0$, 
where $c_0>0$ is a generic constant. 
\end{thm}

Consider inferential learning tasks addressed by learning algorithms or procedures based on quantum walks or underdamped Langevin dynamics. As in the standard machine learning framework, we specify a loss function and evaluate the performance of a learning procedure or algorithm by its associated risk value. The relationship between the performances of learning procedures based on quantum walks and underdamped Langevin dynamics can be characterized using the Le Cam deficiency distance.

Specifically, for two data-generating processes, if their Le Cam distance is zero, then every learning algorithm based on one process has a counterpart based on the other with identical performance, and vice versa. If their Le Cam distance is bounded above by an arbitrarily small constant $\delta$, then every learning procedure or algorithm based on one process admits a corresponding procedure based on the other whose performance differs by at most $\delta$. Furthermore, if the Le Cam distance converges to zero as time tends to infinity, then for any learning algorithm based on one of the two processes, there exists a corresponding algorithm based on the other whose performance difference converges to zero. See \cite{lecam1986} \cite{wang2002} \cite{wang2013asymptotic} for more details.

As established in Theorem 1, the asymptotic result (\ref{mho-1}) implies that $\mho(X^q_t(\eta), X^c_t(\epsilon) )$ can be made arbitrarily small for sufficiently large $t$ and sufficiently small 
$\eta$ and $\epsilon$. This result indicates that every learning algorithm based on $X^q_t(\eta)$ admits a counterpart based on $X^c_t(\epsilon)$ whose performance differs by an arbitrarily small amount. More precisely, for any $\delta>0$, 
Theorem 1 guarantees that $\mho(X^q_t(\eta), X^c_t(\epsilon) ) \leq \delta$ for some sufficiently large $t$ and appropriate choices of small $\eta$ and $\epsilon$. Consequently, for any learning procedure or algorithm based on either $X^q_t(\eta)$ or $X^c_t(\epsilon)$, there exists a corresponding procedure based on the other whose performance gap is no greater than $\delta$. Also, the result can be easily generalized to multiple samples.

Although quantum walks and underdamped Langevin dynamics were developed independently, this asymptotic equivalence establishes a rigorous connection between these previously unrelated frameworks and offers new insight into learning algorithms or procedures 
based on quantum walks and underdamped Langevin dynamics. Since quantum walks and underdamped Langevin dynamics play central roles in the development of fast quantum algorithms in quantum computation and accelerated gradient-based algorithms in classical computation, respectively, our results may stimulate further research into quantum speedup and classical acceleration, as well as deeper exploration of their intrinsic relationships and underlying mechanisms. Additional highlights and details related to these topics are presented in Section \ref{SEC-implication} below.

 The result (\ref{n-X-qc}) established in Theorem 1 suggests that randomization is essential for establishing any learning equivalence between quantum walks and underdamped Langevin dynamics. 
 This phenomenon can be explained heuristically as follows.  For the quantum walk, the measurement $X^q_t$ exhibits asymptotic distributional behavior characterized by high-frequency local oscillations superimposed on a slowly varying global component. 
One may approximate underdamped Langevin dynamics by a quantum walk through an appropriate convolution kernel:  
the slowly varying component of the quantum walk can be captured by the underdamped Langevin dynamics, whereas the oscillatory component, whose frequency is proportional to $1/t$, is difficult to approximate asymptotically by any continuous-time diffusion process. In fact, 
(\ref{mho-1}) implies 
\begin{equation} \label{X-qc-epsilon} 
  \limsup_{t \rightarrow \infty} \gimel ( X^q_{t}, X^c_t(\epsilon))  = O(\epsilon^{1/2} ) .  
\end{equation}  
and the proof of (\ref{n-X-qc}) have actually shown   
\begin{equation} \label{X-qc-epsilon-1} 
  \liminf_{t \rightarrow \infty}  \mho (X^q_t, X^c_t(\epsilon) ) = \liminf_{t \rightarrow \infty} \gimel(X^c_t(\epsilon), X^q_t) \geq c_0 - O(\epsilon^{1/2}) .
\end{equation} 
Numerically, we estimate $c_0 \approx 0.3$, 
although the theoretical lower bound derived in the proof is much smaller. (\ref{X-qc-epsilon}) indicates 
underdamped Langevin dynamics can be approximated by a quantum walk through transformations induced by Markov kernels, whereas (\ref{X-qc-epsilon-1}) shows the reverse approximation is not possible. The proof of (\ref{X-qc-epsilon-1}) further reveals that the limitation arises from 
the high-frequency oscillatory component of the quantum walk.  Consequently, the introduction of randomization via $X^q_t(\eta)$ appears to be essential for establishing learning equivalence between quantum walks and underdamped Langevin dynamics. 
Moreover, since the randomized process $X^q_t(\eta)$ approaches $X^q_t$ as $\eta \rightarrow 0$.  
We may allow $\eta$ to depend on $t$, provided that $\eta t \rightarrow \infty$. That is, $\eta$ must approach zero at a rate much slower than 
$1/t$ to ensure that randomization persists asymptotically. Otherwise, if $\eta$ decreases sufficiently fast so that $[\eta t] <1$, then $X^q_t(\eta)$ coincides with $X^q_t$, and no randomization occurs. 
 
The introduction of $\epsilon$ in our formulations is motivated by intrinsic technical considerations.  
Specifically, the pdf $\pi(y;\theta)$ in (\ref{Konno1}) diverges as $y$ approaches the origin, a singularity that arises from the asymptotic behavior of the quantum walk $X^q_t$. 
This divergence raises concerns about whether the corresponding underdamped Langevin dynamics (\ref{underdamped}) is well defined. 
As $\epsilon \rightarrow 0$, the regularized pdf $\pi_\epsilon(y;\theta)$ converges to $\pi(y;\theta)$, which might allow us to define $X^c_t$ as a suitable limit of $X^c_t(\epsilon)$. For instance, one may  define a distribution $\mP^c_{t,\theta}$ as follows: for any $x$ on the real line,  
\[  \mP^c_{t,\theta}((-\infty, x]) = \limsup_{\epsilon \rightarrow 0} \mP^c_{t, \theta, \epsilon}((-\infty, x]) ,  \]
and define $X_t^c$ as an appropriate limit of $X^c_t(\epsilon)$ as $\epsilon \rightarrow 0$, in the sense that $Pr(X_t^c \in A) = \mP^c_{t,\theta}(A)$ for any  Borel set $A$ on the real line. Under this 
definition, $X^c_t$ may be regarded as the underdamped Langevin dynamics (\ref{underdamped})  with the potential function $\Psi(y) = - \log \pi(y; \theta)$. If $\mP^c_{t,\theta}$ and $X^c_t$ are well defined, we can show  that 
\begin{equation} \label{X-qc-eta} 
   \limsup_{t \rightarrow \infty} \mho ( X^q_{t}(\eta), X^c_t)  = O(\eta^{1/2}) , 
 \end{equation} 
 and 
 \begin{equation} \label{X-qc} 
    \limsup_{t \rightarrow \infty} \gimel ( X^q_{t}, X^c_t)  =   0 ,  \mbox{ but }  \liminf_{t \rightarrow \infty} \gimel ( X^c_{t}, X^q_t)  \geq c_0 >0, 
\end{equation}  
where $c_0$ is the same generic constant as in (\ref{n-X-qc}). 
However, the convergence used in defining $\mP^c_{t,\theta}$ and $X^c_t$ 
may not be weak convergence. 
It is therefore of interest to investigate 
whether $X^c_t$ arises as the weak limit of $X^c_t(\epsilon)$ as $\epsilon \rightarrow 0$. 
Owing to the singularity in $\pi(y;\theta)$, establishing stochastic equicontinuity for $X^c_t(\epsilon)$ is challenging, and we conjecture that $X^c_t(\epsilon)$ 
does not admit a weak limit as $\epsilon \rightarrow 0$.

\subsection{Implications of the equivalence result} 
\label{SEC-implication} 

The asymptotic equivalence established in Theorem 1 may have important implications for the acceleration phenomenon in classical computation and quantum speedup in quantum computation. 
Both quantum walks and underdamped Langevin dynamics can achieve quadratically faster mixing times than their respective classical counterparts. In each case, this quadratic improvement is 
linked to, or even enabled by, an underlying lifting mechanism.

On the quantum side, a quantum walk can offer a quadratic speedup in mixing time relative to its classical analogogue---a Markov chain. Moreover, it is possible to lift the classical 
Markov chain so that the resulting lifted classical Markov chain matches quantum walk in terms of mixing time.  
Lifting is a well-established framework for accelerating MCMC algorithms. Specifically, a lifting of a classical Markov chain $\cM$ is a larger classical Markov chain $\check{\cM}$ constructed by replacing each state of the original chain $\cM$ with a collection of states, with transition probabilities defined so as to preserve the structure of $\cM$.  
In particular, the lifted chain is designed so that collapsing (or projecting) the appropriate states of $\check{\cM}$ recovers the original chain $\cM$, and the invariant distribution of $\check{\cM}$
 projects exactly onto that of $\cM$. 
Such lifting schemes can yield a quadratic speedup in mixing time and thereby significantly reduce the computational cost of  MCMC algorithms. 
Notably, the construction of lifted chains introduces a notion of momentum to achieve this acceleration. 
Quantum walks, which correspond to lifted chains in terms of mixing time,  can also provide a quadratic speed up over the original chain. It is therefore of interest to explore any link between the 
quantum quadratic improvement and lifting mechanisms. 
See 
\cite{apers2017}
\cite{lift-chen}
\cite{dervovic2018}
\cite{diaconis2000}
\cite{hayes2010}
\cite{hildebrand2004}
\cite{jung2007} for more details on lifting. 

On the classical side, the concept of acceleration has become central to gradient-based optimization methods. As a second-order ODE, Nesterov's gradient flow provides a conceptual bridge 
between discrete optimization methods and continuous-time dynamics. 
As a stochastic analogue of Nesterov's gradient flow, the underdamped Langevin dynamics is governed by a second-order SDE. 
There is a natural construction of the underdamped Langevin dynamics via a lifting of the standard (or overdamped) Langevin dynamics from a one-dimensional space to a two-dimensional process. The standard Langevin dynamics is described by the following SDE: 
\[  dY_t = - \nabla \Psi(Y_t) dt + \sqrt{ 2 } \, dB_t , \]  
where $(\nabla, \Psi(\cdot), B_t)$ are defined as in (\ref{underdamped}). 
The underdamped Langevin dynamics augments this formulation by introducing an additional momentum variable alongside the traditional position variable, resulting in a two-dimensional SDE, as given in (\ref{underdamped}).  
For more details, see \cite{fan2021variational} \cite{gardiner1985handbook} \cite{lim2023} \cite{mokrov2021large} \cite{pavliotis2014} 
\cite{song2021}
\cite{vempala2019rapid} 
\cite{yan2023gaussian}
\cite{zhang2024}
\cite{zhang2018policy}.

Sampling algorithms based on Langevin dynamics can be viewed as a form of gradient descent for solving optimization problems on the space of probability measures. In this framework, gradient-based Langevin algorithms lie at the interface of sampling and optimization: they are designed to draw samples from invariant distributions while simultaneously solving optimization problems on the Wasserstein-2 space, with the Kullback-Leibler divergence serving as the objective functional. This optimization perspective naturally motivates the study of acceleration in algorithms derived from Langevin dynamics.

The notion of acceleration has played a key role in gradient-based optimization, with Nesterov’s accelerated gradient descent being a prominent example. In the context of Langevin dynamics, an accelerated algorithm refers to a discretization of the underdamped Langevin dynamics (\ref{underdamped}) 
 that achieves a quadratic improvement in convergence speed over the conventional Langevin algorithm, which corresponds to a straightforward discretization of the standard (overdamped) Langevin diffusion. Indeed, if the Brownian motion term is removed from (\ref{underdamped}), the resulting ODE corresponds to Nesterov’s accelerated gradient flow.
 Viewed as an accelerated variant of Langevin dynamics, the underdamped Langevin dynamics shares key structural similarities with Nesterov’s accelerated gradient flow and enables a quadratically faster sampling rate compared with standard Langevin dynamics. 
See \cite{ambrosio2008gradient}
\cite{calogero2010}
\cite{cao2023}
\cite{cheng2018}
\cite{cheng2018a}
\cite{cheng2020}
\cite{chewi2021analysis}
\cite{eberle2019}
\cite{jordan2017extended}
\cite{jordan1998variational} 
\cite{lambert2022variational} 
\cite{ma2021there}
\cite{pavliotis2014} 
\cite{zuo2024} for more details. 

We have demonstrated that quantum walks and underdamped Langevin dynamics share many structural similarities. Just as standard Langevin dynamics can be viewed as a continuous analogue of a Markov chain, both quantum walks and underdamped Langevin dynamics can be linked to lifting schemes that yield quadratic improvements over Markov chains and standard Langevin dynamics, respectively, in terms of mixing times and algorithmic running times. The asymptotic equivalence established between quantum walks and underdamped Langevin dynamics is expected to stimulate further research into quantum speedup and classical acceleration, as well as deeper exploration of intrinsic connections between these phenomena and the shared underlying mechanisms that give rise to them.

\section{Concluding remarks} 

Search algorithms based on quantum walks can achieve quadratic speedups over their classical counterparts, while gradient-adjusted sampling algorithms derived from underdamped Langevin dynamics can quadratically accelerate conventional Langevin algorithms. Since both search and sampling algorithms are designed to address learning tasks, we have investigated the learning relationship between coined quantum walks and underdamped Langevin dynamics.

We established the asymptotic equivalence, in terms of the Le Cam deficiency distance, between underdamped Langevin dynamics and quantum walks with randomization. However, due to the presence of high-frequency oscillations in the distributional behavior of quantum walks' position measurements, we showed that quantum walks without randomization are not asymptotically equivalent to underdamped Langevin dynamics. We further discussed the implications of these equivalence and nonequivalence results for both the computational and inferential aspects of the associated algorithms in machine learning tasks.

Overall, our findings shed new light on the relationship between quantum walks and underdamped Langevin dynamics, as well as on the intrinsic mechanisms underlying quantum speedup and classical gradient acceleration. We hope that these results will stimulate further research on algorithmic development in both quantum and classical computation.

\section*{Acknowledgements}
The research of Yazhen Wang was supported in part by NSF grant DMS-2514240.

\bibliographystyle{abbrv}     
\bibliography{myReferences-ML}

\section*{Appendix: Proofs of Theorem 1 and  results (\ref{X-qc-epsilon})-(\ref{X-qc})} 

\subsection*{A1. Notations, weak convergence, and the proof plan}
Denote by $\mP_{1t}$ the distribution of 
$X^q_t/t$, by $\mP_{1t\eta}$ the distribution of $X^q(\eta)_t/t$, 
and by $\mP_{2t\epsilon}$ the distribution of $X^c_t(\epsilon)/t$. 
That is, $\mP_{1t}$, $\mP_{1t\eta}$, and $\mP_{2t\epsilon}$ are obtained from $\mP^q_{t,\theta}$, 
$\mP^q_{t, \theta,\eta}$, and $\mP^c_{t, \theta,\epsilon}$ by rescaling with a factor $t$, respectively. 
Here and after, for simplicity we drop subscript $\theta$ without confusion and will write for example $\mho(\mP_{1t\eta}, \mP_{2t\epsilon})$ and 
$\gimel(\mP_{1t\eta}, \mP_{2t\epsilon})$ for $\mho(X^q(\eta)_t/t, X^c_t(\epsilon)/t)$ and $\gimel(X^q_t(\eta)/t, X^c_t(\epsilon)/t)$, respectively. 
 Define a one-to-one transformation $\cT: (-\infty, \infty) \rightarrow (-|u_{11}|, |u_{11}|)$ to be $x=\cT(y) = |u_{11}| e^{-|y|} \mbox{sign}(y)$. Note that the total variation norm is invariant with respect to any one-to-one transformation (e.g. scaling and $\cT$). 
Then we have 
$\mho(X^q(\eta)_t, X^c_t(\epsilon)) = \mho(\mP_{1t\eta}, \mP_{2t\epsilon})$. As $\mho(\mP_{1t\eta}, \mP_{2t\epsilon}) = \max\{ \gimel(\mP_{1t\eta}, \mP_{2t\epsilon}), \gimel(\mP_{2t\epsilon}, \mP_{1t\eta}) \}$, we will show below that as $t \rightarrow \infty$, the limits of $\gimel(\mP_{1t\eta}, \mP_{2t\epsilon})$ and $\gimel(\mP_{2t\epsilon}, \mP_{1t\eta})$ have the bounds given in the theorem. 
 
Let $p(k,t; \theta)$ be the probability function of $X^q_t$. 
Then by definitions we have for any $y \in \bR$,
\begin{equation} \label{equ-cdf}
 \mP_{1t}((-\infty, y]) = 
   \sum_{k \leq t y} p(k, t; \theta) . 
 \end{equation}
The limiting distribution for 
$X^q_t$ as $t \rightarrow \infty$ has been established by theorem 3 in \cite{konno2002}, theorem 1 in \cite{konno2005}, and theorem 1 in \cite{grimmett2004} as follows,  
\begin{equation}\label{limit-QW} 
   \frac{ X^q_t}{t} \Rightarrow \mQ \mbox{ as } t \rightarrow \infty, 
\end{equation} 
where $\Rightarrow$ denotes weak convergence, 
and $\mQ$ has a pdf $f(x; \theta)$ known as the Konno density function given by 
\begin{eqnarray} \label{Konno}
 f(x; \theta) &=& \frac{ \sqrt{ 1- |u_{11}|^2} } {\pi (1 - x^2) \sqrt{ |u_{11}|^2 - x^2} } \nonumber \\
 &&  \left\{ 1 - \left( |a_0|^2  - |a_1|^2 + \frac{ u_{11} a_0 \overline{ u_{12} a_1} + \overline{u_{11} a_0} u_{12} a_1} {|u_{11}|^2} \right)  x \right\} 
\end{eqnarray}
for $x \in (-|u_{11}|, |u_{11}|)$ and  $f(x; \theta) = 0$ otherwise. 
The weak convergence in (\ref{limit-QW}) was proved by showing that the characteristic function of $X^q_t/t$ 
converges to the characteristic function of $\mQ$ as $t \rightarrow \infty$ (see 
\cite{grimmett2004} \cite{kempe2003} \cite{konno2002} \cite{konno2005} \cite{konno2008} \cite{venegas2012}). 
Note that first, $\mQ$ is absolutely continuous with the pdf $f(x; \theta)$; second, the quantum walk at the time $t$ can move at most $t$ steps to the right or left of the starting position, and thus 
$X^q_t/t$ is bounded below and above. Hence, the weak convergence implies 
\[  \lim_{t \rightarrow \infty}     \sum_{k \leq t y} p(k, t; \theta) 
= \mQ( (-\infty, y]) \mbox{ uniformly in $y \in \bR$},  \] 
as well as the convergence of their corresponding moments and moment generating functions. As the probability function $X^q_t(\eta)$ is an average of that for $X^q_t$, by  
Skorokhod's representation theorem we can easily show shat for any $\eta>0$, 
\begin{equation}\label{limit-QW1} 
   \frac{ X^q_t(\eta)}{t} \Rightarrow \mQ \mbox{ as } t \rightarrow \infty .
\end{equation} 

It is easy to see that the Konno density $f(x;\theta)$ is related to the pdf $\pi(y;\theta)$ in (\ref{Konno1}) through the transformation $x=\cT(y)$. 
Define $f_\epsilon(x;\theta) = f(x; \theta)$ for $x \in [ - |u_{11}|(1- \epsilon), |u_{11}| (1-\epsilon)]$, 
$f(x;\theta)= \varkappa_\epsilon$ for $x \in (- |u_{11}|, - |u_{11}| (1- \epsilon)) \cup ( |u_{11}|(1-\epsilon), |u_{11}|)$, and  $f(x; \theta) = 0$ otherwise, where 
$\varkappa_\epsilon$ is given in (\ref{varkappa}) and has the following expression: 
\begin{eqnarray}  \label{varkappa1}
  \varkappa_\epsilon &=& \frac{1}{2 \epsilon|u_{11}|} \left[ 1 -  \int_{-|u_{11}| (1- \epsilon)}^{|u_{11}| (1 - \epsilon)} f(x; \theta) dx \right]  \nonumber \\
  &=&  \frac{ \sqrt{ 1- |u_{11}|^2} } {  \pi \epsilon}  \int_{ 1-\epsilon}^{1}  \frac{ 1 } {(1 - |u_{11}|^2 z^2) \sqrt{ 1 -z^2} }   dz \nonumber \\
    &=& \frac{\sqrt{2}}{\pi \sqrt{\epsilon}  \sqrt{ 1- |u_{11}|^2} }  + O(\sqrt{\epsilon}). 
\end{eqnarray}

\subsection*{A2. Total variation norm result for the Langevin dynamics}

Denote by $\rho_\epsilon(y,v,t)$ the pdf of the underdamped Langevin process $(Y^c_t, V^c_t)$ governed by (\ref{underdamped}) with potential $\Psi_\epsilon (y)$ 
and by $\pi_\epsilon(y,v; \theta)$ its invariant distribution. 
The $\rho_\epsilon(y,v,t)$ satisfies the the Fokker-Planck equation of the following form,
\[   \frac{\partial \rho_\epsilon}{\partial t} = - v \cdot \nabla_y \rho_\epsilon + \nabla_y \Psi_\epsilon \cdot \nabla_v \rho_\epsilon  + \nabla_v \cdot (v \rho_\epsilon) + \Delta_v \rho_\epsilon, \;\; \rho_\epsilon(y, v, 0) = \rho_{\epsilon,0}(y,v), \]
where $\rho_{\epsilon,0}$ is the initial pdf. 
An application of theorem 2.6 and proposition 4.2(i) 
of \cite{stramer1997} leads to that $(Y^c_t, V^c_t)$ is an irreducible T-process with respect to the Lebesgue measure and positive Harris recurrent with invariant measure $\pi_\epsilon(y,v;\theta)$, and thus theorem 6.1 of \cite{meyn1993} indicates that it is ergodic---namely, as $t \rightarrow \infty$, the distribution $\rho_\epsilon(y,v,t)$ converges in the total variation norm to $\pi_\epsilon(y,v;\theta) \propto \exp[ - \Psi_\epsilon(y) - v^2/2]$, where the conditions are verified 
as follows:
\[ \mbox{rank}\left( \begin{pmatrix} 0 \\ 1 \end{pmatrix} , \begin{pmatrix} 0 & 1 \\ 0 & 1\end{pmatrix} \begin{pmatrix} 0 \\  1 \end{pmatrix} 
\right) =2 ; \]

\begin{eqnarray*}
 \nabla \Psi_\epsilon(y)  &=& \mbox{sign}(y) \left[  -1 + \frac{2} { 1 - |u_{11}|^2 e^{-2 |y|} }  + \frac{1}{1 - e^{-2|y|}}   -  \frac{1}{ 1 - \varpi e^{-|y|} } \right] \\
 && \rightarrow \mp 1   \mbox{ as } y \rightarrow \pm \infty ;  
 \end{eqnarray*}
 and
\begin{eqnarray*}
&&  2 \left( \begin{pmatrix} 0 & 0 \\ 0 & 1\end{pmatrix} \left( \begin{array}{c} y \\ v\end{array} \right) , \left( \begin{array}{c} v \\ -v - \nabla \Psi_\epsilon(y) \end{array}\right)  \right) + tr\left( \begin{pmatrix} 0 & 0 \\ 0 & 2\end{pmatrix} \begin{pmatrix} 0 & 0 \\ 0 & 1\end{pmatrix} \right) \\
&& = - 2 v^2 - 2 v \nabla \Psi_\epsilon(y) + 2, 
\end{eqnarray*}
which is bounded by $-v^2$ for large $|y|$ and $|v|$ 
(also see the proof of theorem 2.1 in \cite{roberts1996}). Here 
$\varpi=\left( |a_0|^2  - |a_1|^2 + \frac{ u_{11} a_0 \overline{ u_{12} a_1} + \overline{u_{11} a_0} u_{12} a_1} {|u_{11}|^2} \right)  |u_{11}|$. 

Denote by $\rho_\epsilon(y,t)$ the marginal pdf of $\rho_\epsilon(y,v,t)$ with respect to $y$. Note that 
 $\pi_\epsilon(y;\theta)$ [defined in (\ref{Konno1})] is the marginal pdf of $\pi_\epsilon(y,v;\theta)$ with respect to $y$. 
Then we have 
$\| \rho_\epsilon(y,t) - \pi_\epsilon(y;\theta) \|_{TV} \leq \| \rho_\epsilon(y,v,t) - \pi_\epsilon(y,v;\theta) \|_{TV}
\rightarrow 0$ as $t \rightarrow \infty$. Note that the transformation 
from $Y^c_t$ to $X^c_t(\epsilon)$ is one-to-one, and the distribution $\mP_{2t\epsilon}$ and the pdf $\pi_\epsilon(y;\theta)$ are obtained from the pdfs $\rho_\epsilon(y;t)$ and $f_\epsilon(x; \theta)$ through the transformation $\cT$ plus a scaling factor $t$, respectively. 
Since the total variation norm is invariant under the transformation, $\| \rho_\epsilon(y,t) - \pi_\epsilon(y;\theta) \|_{TV} =\| \mP_{2t\epsilon} - \mQ_\epsilon \|_{TV}$, 
where $\mQ_\epsilon$ denotes the distribution with pdf $f_\epsilon(x;\theta)$.  Therefore, we have that for each $\epsilon$, 
\[  \| \mP_{2t\epsilon} - \mQ_\epsilon \|_{TV} \rightarrow 0 \mbox{ as } t \rightarrow \infty. \]
The total variation distance between $\mQ$ with pdf $f(x;\theta)$ and $\mQ_\epsilon$ with pdf $f_\epsilon(x;\theta)$ has the following bound,  
\begin{eqnarray*}
&&  \| \mQ - \mQ_\epsilon \|_{TV} =\frac{1}{2} \int | f(x;\theta) - f_\epsilon(x;\theta) | dx \\
&& =\frac{1}{2} \int_{ x \in (- |u_{11}|, - |u_{11}| (1- \epsilon)) \cup ( |u_{11}|(1-\epsilon), |u_{11}|) }  | f(x;\theta) - \kappa_\epsilon | dx \\
&& \leq \frac{1}{2} \int_{ x \in (- |u_{11}|, - |u_{11}| (1- \epsilon)) \cup ( |u_{11}|(1-\epsilon), |u_{11}|) }   [ f(x;\theta) + \kappa_\epsilon ]  dx \\
&& = 2 \epsilon |u_{11}|  \varkappa_\epsilon  =   \frac{2\sqrt{2 \epsilon} |u_{11}| }{ \pi \sqrt{ 1 - |u_{11}|^2} }  + O(\epsilon^{3/2})  \\
&& \leq  \frac{2\sqrt{2 \epsilon} }{ \pi \tan \kappa } + O(\epsilon^{3/2})  .
\end{eqnarray*}
Putting the above two results together, we obtain 
\begin{eqnarray} \label{equ-P2Q} 
 && \limsup_{t \rightarrow \infty}   \| \mP_{2t\epsilon} - \mQ\|_{TV}  \leq  \limsup_{t \rightarrow \infty}  \left[  \| \mP_{2t\epsilon} - \mQ_\epsilon \|_{TV} +     \| \mQ_\epsilon - \mQ\|_{TV}
      \right] \nonumber \\
  && \leq  \frac{2\sqrt{2 \epsilon} }{ \pi \tan \kappa }   + O(\epsilon^{3/2}) .
\end{eqnarray} 

\subsection*{A3. Total variation norm result for quantum walk} 

Recall $r = [\eta t]$. $\mP^q_{t,\theta,\eta}$ has a probability function 
\begin{eqnarray*}
&& \sum_{j=-r+1}^{r} Pr(X^q_t + j = y) \frac{1}{2r} = \frac{1}{2r} \sum_{j=-r+1}^{r} p(y-j, t; \theta)  \\
&& = \frac{1}{2r} \sum_{k=y -r}^{y+r-1} p(k,t;\theta) . 
\end{eqnarray*} 

In the randomized scheme for defining $X^q_t(\eta)$, we add an independent uniform random variable $U_j$ on $(-1/2, 1/2]$ to $X^q_t+j$, $j=0, \pm1, \cdots, \pm (r-1), r$, 
and denote by $\mP^q_{*t,\theta,\eta}$ the distribution of the resulting random variable. That is, $\mP^q_{*t,\theta,\eta}$ is the convolution distribution of $\mP^q_{t,\theta,\eta}$ and the $2r$ independent uniform distributions on $(-1/2, 1/2]$. Then $\mP^q_{*t,\theta, \eta}$ is an absolutely continuous distribution with its cdf and pdf as follows: the cdf 
of $\mP^q_{*t,\theta, \eta}$ is given by 
\begin{eqnarray*}
&& \sum_{j=-r+1}^r Pr(X^q_t + j + U_j \leq y) \frac{1}{2r} = \frac{1}{2r} \sum_{j=-r+1}^r Pr(U_j \leq y - j - X^q_t) \\
&& = \frac{1}{2r} \sum_{j=-r+1}^r \sum_\ell p(\ell,t;\theta) Pr(U_j \leq y - j - \ell) \\
&& = \frac{1}{2r} \sum_{j=-r+1}^r p([y]-j, t; \theta) Pr(U_j \leq y - [y]) \\
&& = \frac{1}{2r} \sum_{k=[y]-r}^{[y]+r-1} p(k, t; \theta) Pr(U_j \leq y - [y]) , 
\end{eqnarray*}
where the third equality is due to the fact that as $U_j$ is a uniform distribution on $(-1/2,1/2]$, $y -j -\ell \in (-1/2, 1/2]$ and thus $0=[y-k-\ell] = [y] - k - \ell$; 
and thus its pdf is equal to 
\begin{eqnarray*}
&& \frac{1}{2r} \sum_{k=[y] -r}^{[y]+ r-1} p(k, t;\theta) . \\
\end{eqnarray*}
Denote by $\mP^*_{1t\eta}$ the distribution of $\mP^q_{*t,\theta,\eta}$ obtained through a transformation divided by $t$. 
That is, in the above adding uniform random variable proeedure, we replace $(X^q_t, X^q_t+j, U_j)$ by $(X^q_t/t, (X^q_t+j)/t, U_j/t)$, respectively. 
Note that $\mP_{1t\eta}$ is the distribution obtained from $\mP_{t,\theta,\eta}$ through the same transformation divided by $t$. 
Then $\mP_{1t\eta}$ has a probability function 
\begin{eqnarray*}
&& p_{1t\eta}(x)= \frac{1}{2[\eta t]} \sum_{k=[tx] -r}^{[tx]+ r-1} p(k,t;\theta) , 
\end{eqnarray*}
and $\mP^*_{1t\eta}$ has a pdf 
\begin{eqnarray*}
&& p_{*1t\eta}(x) = \frac{t}{2 [\eta t]} \sum_k p(k,t;\theta) 1\{ -\eta < ( [tx] - k)/t \leq \eta\}  \\
&& = \frac{\eta t}{[\eta t]} \sum_k p(k,t;\theta) \frac{1}{2\eta} 1\{ -\eta < ( [tx] - k)/t \leq \eta\} \\
&& =  \sum_k p(k,t;\theta) \frac{1}{2\eta} 1\{ -\eta < ( [tx] - k)/t \leq \eta\} + O((\eta^2 t)^{-1}), 
\end{eqnarray*}
where the last equality is due to 
\[   \frac{t}{2[\eta t]} = \frac{1}{2 \eta} + O((\eta^2 t)^{-1}) . \] 

Alternatively, we add an independent uniform random variable on $(-\eta, \eta]$ to $X^q_t/t$, 
and denote by $\check{\mP}^*_{1t\eta}$ the distribution of the resulting random variable. That is, $\check{\mP}^*_{1t\eta}$ is the convolution distribution of $\mP_{1t}$ 
and the independent uniform distribution on $(-\eta, \eta]$. $\check{\mP}^*_{1t\eta}$ is absolutely continuous and has a pdf
\begin{equation} \label{equ-p*}
 \check{p}_{*1t\eta}(x) = \sum_{ k } p(k, t; \theta) \frac{1}{2 \eta}  1\{-\eta < ([tx]- k)/t \leq \eta\} .
\end{equation}

Note that 
\begin{eqnarray*}
&&  p_{*1t\eta}(x) = \frac{\eta t}{[\eta t]} \check{p}_{*1t\eta}(x) , 
\end{eqnarray*}
and
\begin{eqnarray} \label{P*Pcheck}
&&\|\check{\mP}^*_{1t\eta} -  \mP^*_{1t\eta} \|_{TV} =  \int | \check{p}_{*1t\eta}(x) - p_{*1t\eta}(x) | dx \nonumber \\
&& =  \left |  \frac{ \eta t}{[\eta t]} - 1 \right| \leq \frac{1}{[\eta t]} =O((\eta t)^{-1} ) .
\end{eqnarray}

Define the Hellinger distance $H(\check{\mP}^*_{1t\eta},\mQ)$ by 
\[  H^2(\check{\mP}^*_{1t\eta},\mQ) 
= \int_\bR \left| \sqrt{  \frac{d\check{\mP}^*_{1t\eta}}{dx}  } - \sqrt{  \frac{d\mQ}{dx}   } \right| dx =  2 - 2 E_{\check{\mP}^*_{1t\eta}}  \left[  \sqrt{  \frac{d\mQ}{d\check{\mP}^*_{1t\eta}} } \right] ,
\]
where 
$dx$ represents the Lebesgue measure. A simple probability analysis leads to 
\begin{equation} \label{TVH-ineq} 
 \| \check{\mP}^*_{1t\eta} - \mQ \|_{TV}^2 \leq H^2(\check{\mP}^*_{1t\eta},\mQ) \leq 2 \check{\mP}^*_{1t\eta}(A^c) + E_{\check{\mP}^*_{1t\eta}}\left [ 1_A \log \frac{d\check{\mP}^*_{1t\eta}}{d\mQ} \right ] , 
\end{equation} 
where $A$ is any fixed Borel set, 
and $A^c$ denotes its complement (see \cite{lecam1986} and equations (42)-(43) in \cite{wang2013asymptotic}). 

For a quantum walk with unitary matrix $\bU$ given in (\ref{equ-U}), $\phi=\pi/4$  corresponds to a general unbiased walk, and the other values of $\phi$ indicates 
biased walks. General unbiased walks have exactly the same probabilistic behavior as the Hadamard walk. The parameter $\phi$ labels a family of walks, and every element of the family gives rise to an identical probability distribution for the same initial condition. Biased walks have probabilistic behaviors similar to the Hadamard walk with some adjustment  (such as using the biased factor $\cos \phi$ instead of $1/\sqrt{2}$) by the biased parameter $\phi$ (see \cite{ambainis2001} \cite{grimmett2004} \cite{nayak2000}). 
For simplicity, we give arguments for the Hadamard walk, with the initiate state $(a_0, a_1) = (1,0)$ and the unitary matrix $\bU$ equal to the Hadamard operator given by (\ref{Hadamard-operator}). 

Let $\zeta$ be a small positive number to be chosen later. Let $A_\zeta = (-1/\sqrt{2} +\zeta, 1/\sqrt{2} - \zeta)$. 
Given $x \in  A_\zeta$, 
for any $t$, we choose an integer $\ell_x=[x t]$, where $[\cdot]$ denotes the nearest integer value. 
Note that $ x - 1/(2t) <\ell_x /t \leq x + 1/(2t)$ and $\ell_x /t \rightarrow x$ as $ t \rightarrow \infty$.  
First we will take $A$ to be $A_\zeta$ in the inequality (\ref{TVH-ineq}). Second, we have 
\begin{equation} \label{equ-RD} 
 \frac{d\check{\mP}^*_{1t\eta}}{d\mQ} = \frac{\check{p}_{*1t\eta}(x)}{f(x;\theta)} ,  
\end{equation} 
where $\check{p}_{*1t\eta}(x)$ denotes the pdf of $\check{\mP}^*_{1t\eta}$ given by (\ref{equ-p*}). 
It is easy to see that $f(x; \theta)$ on $A_\zeta$ is bounded below by $f(0;\theta)$ and above by $\max\{ f(-1/\sqrt{2} + \zeta; \theta),  f(1/\sqrt{2} - \zeta; \theta)\}$. 
Using theorem 2 in \cite{ambainis2001} and equation (8) in \cite{nayak2000}, we can show that when $t$ is large, for $x\in A_\zeta$, $t p([tx], t; \theta)$ is between $ 2 f(x;\theta)$ and $\zeta f(x;\theta)$.
Moreover, as shown in \cite{ambainis2001} and \cite{nayak2000}, we can decompose $t p([tx], t; \theta)$ into slow and fast components, where the slow component is a slowly varying, non-ossillating function to describe the global property of $t p([tx], t; \theta)$, and the fast component is the remaining quickly oscillating function. Viewing as a function of $x$, the slow component is a pdf  equal to $f(x, \theta)$. Denote by $g(x, t; \theta)$ the fast component. Then for $x \in A_\zeta$, 
\begin{equation} \label{p-decompose}
    t p([tx], t; \theta) = f([tx]/t, \theta) + g([tx]/t, t; \theta) , 
    \end{equation} 
where 
\begin{eqnarray}
&& g(x, t; \theta) = f(x;\theta) H(x,t) + O(t^{-1}), \label{g-H} \\
&& H(x,t) = \sqrt{2} x \sin( 2[\omega_{\alpha_x} + x \alpha_x ] t - \beta_x), \label{equ-H}  \\
&& \omega_{\alpha} = \sin^{-1} \left( \frac{\sin \alpha }{\sqrt{2}} \right), \; \cos \alpha_x = \frac{x}{\sqrt{1-x^2}}, \; \cos \beta_x = \frac{2 x -1}{\sqrt{2} (1-x)} . \label{a-b-o} 
\end{eqnarray} 
According to (\ref{p-decompose})-(\ref{a-b-o}), for $x \in A_\zeta$ we write $\check{p}_{*1t\eta}(x)$ in (\ref{equ-p*}) as the sum of $I_1(x)$ and $I_2(x)$, where   
\begin{eqnarray*}
&& I_1(x)= \sum_{u_k=k/t}  \frac{1}{t} f(u_k; \theta) \frac{1}{2 \eta}  1\{-\eta < x - u_k < \eta\}  \\
&& = \frac{1}{2\eta} \int_{x -\eta}^{x+\eta} f(u;\theta) du + O(t^{-1}) \\
&&
= f(x;\theta) + O(\eta) + O(t^{-1}), \\
&& I_2(x) = \sum_{u_k=k/t} \frac{1}{t} g(u_k, t;\theta)  \frac{1}{2 \eta}  1\{-\eta < x-u_k< \eta\} \\
&& = \sum_{u_k=k/t} \frac{1}{t} f(u_k; \theta) H(u_k,t) \frac{1}{2 \eta}  1\{-\eta < x - u_k < \eta\} + O(\eta t^{-1})\\
&& = \frac{1}{2\eta} \int_{x-\eta}^{x+\eta} f(u;\theta) H(u,t)  du + O(t^{-1})  + O(\eta t^{-1})  \\
&& = O(\eta + (\eta t)^{-1} )  + O(t^{-1})  + O(\eta t^{-1}) . 
\end{eqnarray*}
Here the last equality in the above analysis of $I_2(x)$ is due to the following fact: 
\begin{eqnarray*}
&& \frac{1}{2\eta} \int_{x-\eta}^{x+\eta} f(u;\theta) H(u,t)  du = 
 \frac{1}{2 \eta} \int_{x-\eta}^{x+\eta} f(u;\theta) \sqrt{2} u \sin \left( 2[\omega_{\alpha_u} + u \alpha_u ] t - \beta_u \right) du \\
&& = \frac{1}{\sqrt{2} \eta} \int_{x-\eta}^{x+\eta} f(x;\theta) x \sin \left( 2[\omega_{\alpha_u} + u \alpha_u ] t - \beta_x \right) du + O(\eta) \\ 
&&= \frac{f(x;\theta) x}{\sqrt{2} \eta}  \left(  \frac{ d [\omega_{\alpha_x} + x \alpha_x]} { dx} \right)^{-1}   \frac{1}{2 t }  \{- \cos \left( 2[\omega_{\alpha_u} + u \alpha_u ] t   - \beta_x\right) \} |^{u=x+\eta}_{u=x-\eta} + O(\eta)  \\
&& = O( (\eta t)^{-1} + \eta) . 
\end{eqnarray*}
Using the obtained expressions for $I_1(x)$ and $I_2(x)$ and $\check{p}_{*1t\eta}(x)=I_1(x) + I_2(x)$, we get that for $x \in A_\zeta$,
\begin{equation}  \label{p*t-epsilon}
  \check{p}_{*1t\eta}(x)  =  f(x; \theta)  +O(\eta + (\eta t)^{-1} ) .  
\end{equation}

Combining (\ref{equ-RD}) and (\ref{p*t-epsilon}) together we obtain that for $x \in A_\zeta$, 
\[  \log \frac{d\check{\mP}^*_{1t\eta}}{d\mQ} = \log \left[ \frac{ f(x; \theta)  +O(\eta + (\eta t)^{-1} ) 
 }{f(x;\theta)} \right]  \leq \frac{O(\eta + (\eta t)^{-1} ) } {f(x;\theta)} , \]
 and thus, applying the dominated convergence theorem, we have 
 \begin{equation} \label{RD-converge1} 
  \limsup_{t \rightarrow \infty}  E_{\check{\mP}^*_{1t\eta}} \left [  1_{A_\zeta}  \log \frac{d\check{\mP}^*_{1t\eta}}{d\mQ} \right] = O(\eta).
 \end{equation}

The event $X^q_t \in t A_\zeta = ((-1/\sqrt{2} +\zeta)t , (1/\sqrt{2} -\zeta) t)$ means that 
 quantum walk 
 has a measurement outcome at time $t$ falling inside of the interval  
$((-1/\sqrt{2} +\zeta)t , (1/\sqrt{2} -\zeta) t)$. 
From theorem 1 in  \cite{ambainis2001} we have that as $t \rightarrow \infty$, $p([xt], t; \theta) = O(\varrho^{t})$ for $x \in (-1, -1/\sqrt{2}) \cup (1/\sqrt{2}, 1)$, where $\varrho \in (0,1)$ is a fixed constant. Using expressions in theorem 2 of  \cite{ambainis2001} for the limit of $p(k,t;\theta)$ as $t\rightarrow \infty$ we can directly calculate 
\[ \sum_{x=k/t, k \in  [(-1/\sqrt{2} + \zeta) t , (1/\sqrt{2} - \zeta)  t]} p(x, t; \theta) = 1 - \frac{2 \zeta}{\pi} - O(t^{-1}) , \] 
and 
\[ \sum_{x=k/t, k \in  [-t/\sqrt{2}, (-1/\sqrt{2} +\zeta) t]  \cup [(1/\sqrt{2} -\zeta) t, t/\sqrt{2}] } p(x, t; \theta) = \frac{2\zeta}{\pi} + O(t^{-1})  . \]
In fact, the above results were given in discussions after theorem 2 of  \cite{ambainis2001} and theorem 4 of \cite{venegas2012} as well as appendix C of 
\cite{nayak2000}. Therefore, we have 
\begin{eqnarray} \label{equ-PA} 
 \check{\mP}^*_{1t\eta}(A^c_\zeta) &\leq& \mbox{Prob}(X_t^q/t \in A^c_\zeta) =\mP_{1t}(A^c_\zeta)=  \sum_{x=k/t, k \in  (-t, -t/\sqrt{2})  \cup (t/\sqrt{2}, t) } p(x, t; \theta) \nonumber \\
&& + \sum_{x=k/t, k \in  [-t/\sqrt{2}, (-1/\sqrt{2} + \zeta)t]  \cup [(1/\sqrt{2} - \zeta) t, t/\sqrt{2}] } p(x, t; \theta) \nonumber \\
&& = O(t \varrho^{t})  + \frac{2\zeta}{\pi} + O(t^{-1}) .
\end{eqnarray}
 
Plugging (\ref{RD-converge1}) and  (\ref{equ-PA}) into (\ref{TVH-ineq}), we conclude that 
 \begin{eqnarray} \label{RD-converge2} 
 &&  
  \limsup_{t \rightarrow \infty}  \| \check{\mP}^*_{1t\eta}- \mQ \|_{TV}^2 
    \leq  2 \limsup_{t \rightarrow \infty}  \left[ 
 O(t \varrho^{t})  + \frac{2\zeta}{\pi} + O(t^{-1}) \right] + O(\eta) \nonumber \\
 && = 
  \frac{4\zeta}{\pi}  +  O(\eta) .
  \end{eqnarray} 
Since the left-hand side of (\ref{RD-converge2}) is free of $\zeta$, we let $\zeta$ go to zero and obtain 
\begin{equation} \label{equ-P*Q} 
  \limsup_{t \rightarrow \infty}  \| \check{\mP}^*_{1t\eta}- \mQ \|_{TV} = O(\eta^{1/2}) . 
\end{equation} 

\subsection*{A4. Result (\ref{mho-1}) in Theorem 1  for Le Cam distance between Langevin dynamics and quantum walk} 

Putting together (\ref{equ-P2Q}), (\ref{P*Pcheck}), and (\ref{equ-P*Q})  we conclude 
\begin{eqnarray} \label{equ-P*1P2} 
\limsup_{t \rightarrow \infty}  \| \mP^*_{1t\eta}- \mP_{2t\epsilon}\|_{TV} &\leq& \limsup_{t \rightarrow \infty} [ \| \mP^*_{1t\eta} - \check{\mP}^*_{1t\eta} \|_{TV} + 
\| \check{\mP}^*_{1t\eta}- \mQ \|_{TV} \nonumber \\ 
&& + \| \mP_{2t\epsilon} - \mQ \|_{TV} ] \nonumber \\
&\leq&   \frac{2\sqrt{2 \epsilon} }{ \pi \tan \kappa }  +  O(\epsilon^{3/2} + \eta^{1/2}), 
\end{eqnarray} 
which holds for any $\theta \in \Theta$. Note that $\varphi$ belongs to  a closed set on the unit sphere in $\mC^2$, 
$\bU$ lies in the compact set $\{|u_{11}|^2 + |u_{12}|^2=|u_{21}|^2+|u_{22}|^2=1\}$, and $\Theta$ is compact. Also, 
$\| \mP^*_{1t\eta}- \mP_{2t\epsilon}\|_{TV}$ is continuous in $\theta$, as $\mP_{1t}$, $\mP^*_{1t\eta}$ and $\mP_{2t\epsilon}$ continuously depend on $\theta$. Thus, 
\begin{equation} \label{equ-P*1P2-sup} 
 \limsup_{t \rightarrow \infty} \sup_{\theta \in \Theta} \| \mP^*_{1t\eta}- \mP_{2t\epsilon}\|_{TV} \leq \frac{2\sqrt{2 \epsilon} }{ \pi \tan \kappa }  + O(\epsilon^{3/2} + \eta^{1/2}) . 
\end{equation} 
Finally, using the fact that the convolution of $\mP_{1t\eta}$ and uniform distributions is a particular Markov kernel, we obtain 
\begin{eqnarray} \label{gimel12}
  \limsup_{t \rightarrow \infty} \gimel(\mP_{1t\eta}, \mP_{2t\epsilon}) &\leq& \limsup_{t \rightarrow \infty}\sup_{\theta \in \Theta} \| \mP^*_{1t\eta}- \mP_{2t\epsilon}\|_{TV} \nonumber \\
&\leq& \frac{2\sqrt{2 \epsilon} }{ \pi \tan \kappa }  + O(\epsilon^{3/2} + \eta^{1/2}) . 
\end{eqnarray}

For the case of $\gimel(\mP_{2t\epsilon}, \mP_{1t\eta})$, we reverse the above convolution scheme by discretizing $\mP_{2t\epsilon}$ in order to compare it with $\mP_{1t\eta}$. Specifically, define $\mP^*_{2t\epsilon}$ to be the discrete distribution taking values $k/t$, $k=0, \pm 1, \pm2, \cdots$,
with probability function at $k/t$ equal to $\mP_{2t\epsilon}(( \frac{k}{t} - \frac{1}{2t}, \frac{k}{t} + \frac{1}{2t} ])$. Also, we discretize $\mP^{*}_{1t\eta}$. 
Let $\mP^{**}_{1t\eta}$ be the discrete distribution taking values $k/t$, $k=0, \pm 1, \pm2, \dots$, with probability function at $k/t$ equal to 
$\mP^{*}_{1t\eta}(( \frac{k}{t} - \frac{1}{2t}, \frac{k}{t} + \frac{1}{2t} ])$. Note that $r = [ \eta t]$. 
By (\ref{equ-p*}) we have for $x=\ell_x/t$ or $\ell_x = x t$, 
\begin{eqnarray*}
&& 
\mP^{*}_{1t\eta} \left(\left( \frac{k}{t} - \frac{1}{2t}, \frac{k}{t} + \frac{1}{2t} \right] \right) = \int_{x-1/(2t)}^{x+1/(2t)} p^*_{1t\eta}(u) du \\
&& = \sum_{ k =\ell_x-r}^{\ell_x+r-1} p(k, t; \theta) \frac{1}{2 [\eta t]}  = 
  \frac{1}{2r} \sum_{ k =\ell_x-r}^{\ell_x+r-1} p(k, t; \theta) = p_{1t\eta}(x) . 
\end{eqnarray*}
Hence, $\mP^{**}_{1t\eta} = \mP_{1t\eta}$, that is, discretizing $\mP^*_{1t\eta}$ recovers $\mP_{1t\eta}$. 

Since $\mP^*_{2t\epsilon}$ and $\mP_{1t\eta}$ are discretized versions of $\mP_{2t\epsilon}$ and $\mP^*_{1t\eta}$, respectively, and the discretization reduces their corresponding total variation distance, we have $\| \mP_{1t\eta} - \mP^*_{2t\epsilon} \|_{TV} \leq \| \mP^*_{1t\eta} - \mP_{2t\epsilon} \|_{TV}$. 
Thus, (\ref{equ-P*1P2}) implies 
\begin{equation}  \label{equ-P1P*2} 
\limsup_{t \rightarrow \infty}  \| \mP_{1t\eta}- \mP^*_{2t\epsilon}\|_{TV} \leq   \frac{2\sqrt{2 \epsilon} }{ \pi \tan \kappa }  +  O(\epsilon^{3/2} + \eta^{1/2}) .
\end{equation}
Again, using the continuity of $\mP_{1t\eta}$ and $\mP^*_{2t\epsilon}$ in $\theta$ and the compactness of $\Theta$, we  
conclude that the limit of $ \sup_{\theta \in \Theta|} \| \mP_{1t\eta} - \mP^*_{2t\epsilon}  \|_{TV}$ has the same upper bound as in (\ref{equ-P1P*2}), 
which in turn implies that the same upper bound holds for the limit of $\gimel(\mP_{2t\epsilon}, \mP_{1t\eta})$, as the discretization is a Markov kernel. 
That is,
\begin{equation} \label{gimel21} 
  \limsup_{t \rightarrow \infty} \gimel(\mP_{2t\epsilon}, \mP_{1t\eta}) \leq \frac{4\sqrt{2 \epsilon} }{ \pi \tan \kappa }  + O(\epsilon^{3/2}+ \eta^{1/2}) . 
\end{equation} 
The first result (\ref{mho-1}) of Theorem 1 is a consequence of (\ref{gimel12}) and (\ref{gimel21}).

\subsection*{A5. Result (\ref{n-X-qc}) in Theorem 1 for Le Cam distance between Langevin dynamics and quantum walk} 

In the definitions of the deficiency $\gimel$ and Le Cam distance $\mho$, we need to take the supremum over the parameter space. Thus, shrinking the parameter space decreases the deficiency $\gimel$ and Le Cam distance $\mho$. To establish a positive lower bound of (\ref{n-X-qc}), we may reduce the parameter space $\Theta$ to a single point set $\Theta_0=\{\theta_0\}$ that corresponds to 
$\gamma_1=\gamma_2=\varsigma_1=0$, $\varsigma_2 = \pi/2$, and  $\vartheta=\phi=\pi/4$ in (\ref{equ-U}) and (\ref{Theta}). In the rest of the theorem  proof we consider the single point parameter space $\Theta_0$ without any 
supremum step in handling $\gimel$ and $\mho$. 

Using the triangle property of the distance $\mho$ and the relationship of $\mho$ and $\gimel$, we have 
\begin{eqnarray} \label{n-X-qc-epsilon} 
  \mho(\mP_{2t\epsilon}, \mP_{1t}) &\geq& \mho(\mQ^*, \mP_{1t}) - \mho(\mQ^*, \mP_{2t\epsilon})  \nonumber \\
  &\geq& \gimel(\mQ^*, \mP_{1t}) -  \mho(\mQ^*, \mQ) - \mho(\mQ, \mP_{2t\epsilon}) , 
  \end{eqnarray} 
 where $\mQ^*$ is defined to be the discrete distribution taking values $k/t$, $k=0, \pm 1, \pm2, \cdots$,
with probability function at $k/t$ equal to $\mQ(( \frac{k}{t} - \frac{1}{2t}, \frac{k}{t} + \frac{1}{2t} ])$. 
 We will analyze the three terms on the right-hand side of (\ref{n-X-qc-epsilon}).

 First, it is easy to see that (\ref{equ-P2Q}) implies 
 \begin{equation} \label{n-Q-P2-epsilon}
\limsup_{t \rightarrow \infty}  \mho(\mQ, \mP_{2t\epsilon}) = O(\epsilon^{1/2}) . 
\end{equation} 

Second, we will show 
\begin{equation} \label{n-Q-Q*} 
 \limsup_{t \rightarrow \infty} \mho(\mQ^*, \mQ) = 0 . 
 \end{equation} 
Since $\mQ^*$ is a discretization of $\mQ$, and the discretization is a Markov kernel, 
we have $\gimel(\mQ, \mQ^*)=0$. To handle $\gimel(\mQ^{*}, \mQ)$, we define $\mQ^{**}$ to be the convolution of $\mQ^*$ and independent uniform distributions 
on $( \frac{k}{t} - \frac{1}{2t}, \frac{k}{t} + \frac{1}{2t} ]$. As the convolution is a Markov kernel, we obtain 
$\gimel(\mQ^{*}, \mQ) \leq \| \mQ^{**} - \mQ\|_{TV}$. So it is enough to establish the zero limit for the total variation distance between $\mQ$ and $\mQ^{**}$. 
Denote by $f^{**}(x;\theta_0)$ the pdf of $\mQ^{**}$. Then 
we have  for $x \in (-1/\sqrt{2}, 1/\sqrt{2})$ and $k=[tx]$, 
\begin{eqnarray} \label{equ-f**}
&& f^{**}(x,\theta_0) = t \mQ\left( \left( \frac{k}{t} - \frac{1}{2t}, \frac{k}{t} + \frac{1}{2t} \right] \right)  \nonumber \\
&& = f(k/t; \theta_0) + O(t^{-1}) = f([tx]/t;\theta_0) + O(t^{-1}) . 
\end{eqnarray}
Here we derive the second equality in (\ref{equ-f**}) by the following two results:
\begin{equation} \label{pQ-2} 
  \mQ\left( \left(  \frac{k}{t} - \frac{1}{2 t} ,  \frac{k}{t} + \frac{1}{2 t} \right] \right) = \int_{k/t-1/(2t)}^{k/t+1/(2t)} f(u; \theta_0) du 
 = \frac{1}{t}  f(\check{x}; \theta_0), 
\end{equation} 
where the last equlaity is due to the mean value theorem, and $\check{x} \in [ k/t - 1/(2t), k/t + 1/(2t)]$, which implies $x - 3/(2t) \leq \check{x} \leq x + 1/(2t)$ and 
\begin{equation} \label{pQ-3}
  f(\check{x}; \theta_0) = f(x;\theta_0) + O(t^{-1}) = f([tx]/t;\theta_0) + O(t^{-1}). 
\end{equation}

Recall $A_\zeta = (-1/\sqrt{2} +\zeta, 1/\sqrt{2} - \zeta)$ for $\zeta>0$. We have for $x \in A_\zeta$, 
\begin{eqnarray} \label{RD-converge4}
 && \frac{d\mQ}{d\mQ^{**}} = 
 \frac{ f(x; \theta_0)}{ f^{**}(x;\theta_0) }  \nonumber \\
 && = \frac{ f(x; \theta_0)}{f([tx]/t;\theta_0)+ O(t^{-1})} \nonumber \\
 && = 1 + O(t^{-1}),
 \end{eqnarray} 
where the second equality is due to (\ref{equ-f**}), and 
the last equality is from 
\[ x - \frac{1}{2t} \leq \frac{[tx]}{t} \leq x + \frac{1}{2t}. \]
Giving the expression of $\frac{d\mQ}{d\mQ^{**}}$ in (\ref{RD-converge4}), and on $A_\zeta$, $\frac{d\mQ}{d\mQ^{**}}$ is bounded. 
Thus, 
(\ref{RD-converge4}) and the dominated convergence theorem indicate 
\begin{equation} \label{RD-converge5}
  \limsup_{t\rightarrow \infty}  E_{\mQ} \left [ 1_{A_\zeta} \log \frac{d\mQ} {d\mQ^{**}}\right ] = 0 . 
\end{equation} 
The same calculation as in (\ref{varkappa1}) leads to 
\begin{equation} \label{equ-QA} 
\mQ(A^c_\zeta) = \int_{A^c_\zeta}  f(x; \theta_0) dx = \int_{|x|>1/\sqrt{2} - \zeta} f(x; \theta_0) dx = \frac{2^{3/4} \zeta^{1/2} }{\pi} + O(\zeta^{3/2}). 
\end{equation} 
Applying the inequality (\ref{TVH-ineq}) to $\mQ$ and $\mQ^{**}$, and using (\ref{RD-converge5})-(\ref{equ-QA}), we get 
\begin{eqnarray*}
&&  \limsup_{t \rightarrow \infty}   \| \mQ - \mQ^{**} \|_{TV}^2  \leq \limsup_{t \rightarrow \infty} \left\{
 2 \mQ(A^c_\zeta) + E_{\mQ}\left [ 1_{A_\zeta} \log \frac{d\mQ}{d\mQ^{**}} \right ] \right\} \\
 && = \frac{2^{2+3/4} \zeta^{1/2} }{\pi} + O(\zeta^{3/2}) . 
 \end{eqnarray*}
Since the left-hand side of the above inequality is free of $\zeta$, we let $\zeta \rightarrow 0$ to obtain 
\[   \limsup_{t \rightarrow \infty}  \| \mQ - \mQ^{**} \|_{TV}  =0  , \]
which competes the proof of (\ref{n-Q-Q*}). 

Third, we derive a lower bound for $\gimel(\mQ^*, \mP_{1t})$. 
Denote by $(\lambda_{jk})$ a stochastic matrix to represent a Markov kernel satisfying $\lambda_{jk} \geq 0$ for all $(j,k)$ and $\sum_{j} \lambda_{jk} =1$ for all $k$. Then we have 
\begin{eqnarray} \label{n-Q*-P1} 
&& \gimel(\mQ^*, \mP_{1t}) = \inf_{(\lambda_{jk})} \sum_k \left | p(k, t; \theta_0) -  \sum_{j} \lambda_{jk} \mQ\left(\left( \frac{j}{t} - \frac{1}{2t}, \frac{j}{t} + \frac{1}{2t} \right] \right) 
 \right | \nonumber \\
&& = \inf_{(\lambda_{jk})} \sum_k \frac{1}{t} \left | f(k/t; \theta_0) [1 + H(k/t,t)] - \sum_j \lambda_{jk} f(j/t;\theta_0) \right|  + O(t^{-1})   \nonumber \\
&& \geq  \inf_{\lambda(y,x)}  \int_{-1/\sqrt{2}}^{1/\sqrt{2}} \left | f(x; \theta_0) [1 + H(x,t)]  -  \int_{-1/\sqrt{2}}^{1/\sqrt{2}} \lambda(y,x) f(y;\theta_0) dy  \right | dx + O(t^{-1})  \nonumber \\
&& \geq \inf_{\lambda(y,x)} \int_{\Omega_{t}} \left | f(x; \theta_0) [1 + H(x,t)]  -  \int_{-1/\sqrt{2}}^{1/\sqrt{2}} \lambda(y,x) f(y;\theta_0) dy  \right | dx + O(t^{-1})   \nonumber \\
&& \geq \int_{\Omega_{t}}  f(x; \theta_0) | H(x,t) | dx + O(t^{-1})  \nonumber \\
&& =  \frac{1}{2} \int^{1/\sqrt{2}}_{-1/\sqrt{2}} f(x; \theta_0) | H(x,t) | dx  + O(t^{-1})  \nonumber \\
&& \geq \frac{1}{2 \pi}  \int^{1/\sqrt{2}}_{-1/\sqrt{2}}  | H(x,t) | dx +O(t^{-1}), 
\end{eqnarray}
where $H(x,t)$ is defined in (\ref{equ-H}), $\lambda(y,x)$ represents a kernel function satisfying $\lambda(y,x)\geq 0$ for all $(y,x)$ and $\int \lambda(y,x) dy =1$ for all $x$, 
$\Omega_{t}=\{x \in (-1/\sqrt{2}, 1/\sqrt{2}): H(x, t) \leq 0\}$, the equality in the first array uses the definition of $\gimel(\mQ^*, \mP_{1t})$ with a single point parameter space $\Theta_0$, 
the equality in the second array is from (\ref{p-decompose}) and (\ref{pQ-2})-(\ref{pQ-3}), 
the inequality in the third array uses the integral approximation and that kernel $\lambda(y,x)$ covers the discrete case of $\lambda_{jk}$, the inequality in the fourth array is because of 
$\Omega_t \subset (-1/\sqrt{2}, 1/\sqrt{2})$, the inequality in the fifth array  is due to the concavity of $f(x;\theta_0)$ which implies 
\[  f(x; \theta_0)  -  \int_{-1/\sqrt{2}}^{1/\sqrt{2}} \lambda(y,x) f(y;\theta_0) dy \leq 0 ,  \]
the equality in the sixth array is attributed to the following results: 
\begin{eqnarray*}
&& 1 = \sum_k p(k,t;\theta_0) = \sum_k \{  f(k/t;\theta_0) [ 1 + H(k/t,t) ]/t + O(t^{-2}) \} \\
&& = \int^{1/\sqrt{2}}_{-1/\sqrt{2}}  f(x;\theta_0) [1 + H(x,t)] dx + O(t^{-1}) \\
&& = 1 + \int f(x;\theta_0) H(x,t) dx + O(t^{-1}) \\
&& 
     = 1 - \int_{\Omega_t} f(x;\theta_0) | H(x,t) | dx + \int_{\Omega_t^c} f(x;\theta_0) | H(x,t)| dx + O(t^{-1}),  \\
&& \int_{\Omega_t} f(x;\theta_0) | H(x,t) | dx =  \int_{\Omega_t^c} f(x;\theta_0) | H(x,t)| dx + O(t^{-1}) \\
&& = \int_{\Omega_t^c} f(x;\theta_0) | H(x,t)| dx + \frac{1}{2} \int f(x;\theta_0) H(x,t)dx + O(t^{-1}) \\
&& = \frac{1}{2} \int f(x;\theta_0) |H(x,t)| dx + O(t^{-1}) ,
\end{eqnarray*}
and the last inequality is from 
\[  f(x;\theta_0) = \frac{1}{\pi (1-x^2) \sqrt{1-2x^2} } \geq \frac{1}{\pi} . \]

We will show below 
\[ \liminf_{t \rightarrow \infty} \int^{1/\sqrt{2}}_{-1/\sqrt{2}} |H(x,t)| dx > 0 . \] 
Define  $x_\ell$, 
$\ell=0, \pm 1, \cdots, \pm L_t$, to be a sequence of all points in the interval $(-1/\sqrt{2}, 1/\sqrt{2})$ that satisfy 
\[    2[\omega_{\alpha_{x_{\ell}}} + x_\ell \alpha_{x_\ell} ] t - \beta_{x_\ell}   = \pi \ell, \]
and we pick a point $\xi_\ell \in (x_\ell, x_{\ell+1})$ to make  
\[  2[\omega_{\alpha_{\xi_{\ell}}} + \xi_\ell \alpha_{\xi_\ell} ] t - \beta_{\xi_\ell}  \in [\pi \ell + \pi/4, \pi \ell + 3\pi/4] . 
\]
Since $x_\ell, \xi_\ell \in (-1/\sqrt{2}, 1/\sqrt{2})$, we conclude that for large $t$, $L_t \propto t$, and  
\begin{eqnarray*}
&&  2[\omega_{\alpha_{x_{\ell+1}}} + x_{\ell+1} \alpha_{x_{\ell+1}} ] t - \beta_{x_{\ell+1}} - 2[\omega_{\alpha_{x_{\ell}}} + x_\ell \alpha_{x_\ell} ] t - \beta_{x_\ell}   = \pi  , \\
&&  2[\omega_{\alpha_{\xi_{\ell}}} + \xi_{\ell} \alpha_{\xi_{\ell}} ] t - \beta_{\xi_{\ell}} - 2[\omega_{\alpha_{x_{\ell}}} + x_\ell \alpha_{x_\ell} ] t - \beta_{x_\ell}   \in [\pi/4, 3\pi/4] ,\\
&& \{ 2[\dot{\omega}_{\alpha_{x_{\ell}}} + x_\ell \dot{\alpha}_{x_\ell} + \alpha_{x_\ell} ] t  - \dot{\beta}_{x_\ell}  \} (x_{\ell+1}-x_{\ell} ) = \pi  + o(1) , \\
&& \{ 2[\dot{\omega}_{\alpha_{x_{\ell}}} + x_\ell \dot{\alpha}_{x_\ell} + \alpha_{x_\ell} ] t  - \dot{\beta}_{x_\ell}  \} (\xi_{\ell}-x_{\ell} )  \in [\pi/4, 3\pi/4] + o(1) , \\
&& x_{\ell+1}-x_{\ell} = \frac{\pi}{ | 2[\dot{\omega}_{\alpha_{x_{\ell}} } + x_\ell \dot{\alpha}_{x_\ell} + \alpha_{x_\ell} ] t  - \dot{\beta}_{x_\ell} | } + o(t^{-1})  , \\
&& \xi_{\ell}-x_{\ell} \in \frac{1 }{  | 2[\dot{\omega}_{\alpha_{x_{\ell}} } + x_\ell \dot{\alpha}_{x_\ell} + \alpha_{x_\ell} ] t  - \dot{\beta}_{x_\ell} | } [\pi/4, 3\pi/4]+ o(t^{-1}) .
\end{eqnarray*}\
Note that the sine function in $H(x,t)$ is equal to $0$ at $x_\ell$ and is either larger than $1/2$ or smaller than $-1/2$ at $\xi_\ell$ (depending on its value is positive or negative),  
and its area between $x_\ell$ and $x_{\ell+1}$ is less than the area of 
a triangle formed by the three points $(x_{\ell}, 0)$, $(x_{\ell+1},0)$, and $(\xi_\ell, \pm 1/2)$ (according to the area above or below the horizontal axis).  
Hence, we obtain 
\begin{eqnarray} \label{n-H-bound} 
&&\int^{1/\sqrt{2}}_{-1/\sqrt{2}}  | H(x,t) | dx \geq \sum_{\ell=-L_t}^{L_t} \frac{|\xi_\ell | }{4} (x_{\ell+1} - x_{\ell}) \nonumber \\
&& = \sum_{\ell=-L_t}^{L_t} \frac{|x_\ell | }{4} \frac{\pi}{ |  2[\dot{\omega}_{\alpha_{x_{\ell}} } + x_\ell \dot{\alpha}_{x_\ell} + \alpha_{x_\ell} ] t  - \dot{\beta}_{x_\ell}  | } + o(1)  \nonumber \\
&& = \int^{1/\sqrt{2}}_{-1/\sqrt{2}} \frac{|x| }{4} \frac{\pi }{ |  2[\dot{\omega}_{\alpha_{x}} + x \dot{\alpha}_{x} + \alpha_{x} ] t  - \dot{\beta}_{x}  | } \frac{ 2L_t+1}{\sqrt{2}} dx + o(1) .
\end{eqnarray}
Putting these results together we conclude 
\begin{eqnarray} \label{n-Q*-P1-bound} 
&& 
    c_0 = \liminf_{t \rightarrow \infty }   \gimel(\mQ^*, \mP_{1t}) \geq
 \frac{1}{2\pi} \liminf_{t \rightarrow \infty }  \int^{1/\sqrt{2}}_{-1/\sqrt{2}}  | H(x,t) | dx \nonumber \\
&& \geq \frac{1}{2\pi}
  \int^{1/\sqrt{2}}_{-1/\sqrt{2}} \frac{|x| }{4} \liminf_{t \rightarrow \infty } \left[ \frac{\pi }{ |  2[\dot{\omega}_{\alpha_{x}} + x \dot{\alpha}_{x} + \alpha_{x} ] t  - \dot{\beta}_{x}  | } 
  \frac{2 L_t +1}{\sqrt{2}}   \right] dx \nonumber \\
&& \geq \frac{1}{8\pi}  \min_{-1 \leq \sqrt{2}x \leq 1}\{  | \dot{\omega}_{\alpha_x} + x \dot{\alpha}_{x} + \alpha_{x}  |  \}
\int^{1/\sqrt{2}}_{-1/\sqrt{2}}  \left| \frac{ x }{ \dot{\omega}_{\alpha_{x}} + x \dot{\alpha}_{x} + \alpha_{x} } \right| dx >0 , \;\;\;\;\;
\end{eqnarray}
where the first inequality is from (\ref{n-Q*-P1}),  the second inequality is due to (\ref{n-H-bound}) and Fatou's lemma, and the last inequality uses the following results: 
\begin{eqnarray*}
&& \liminf_{t \rightarrow \infty } \left[ \frac{\pi t }{ |  2[\dot{\omega}_{\alpha_{x}} + x \dot{\alpha}_{x} + \alpha_{x} ] t  - \dot{\beta}_{x}  | } \right] = \frac{\pi }{ |  2[\dot{\omega}_{\alpha_{x}} + x \dot{\alpha}_{x} + \alpha_{x} ]  | } ,\\
&& 2 L_t + 1 \geq \frac{\sqrt{2}}{\pi} \min \{ | 2[\dot{\omega}_{\alpha_{x_{\ell}} } + x_\ell \dot{\alpha}_{x_\ell} + \alpha_{x_\ell} ] t  - \dot{\beta}_{x_\ell} | : -L_t \leq \ell \leq L_t \} \\
&& \geq \frac{\sqrt{2}}{\pi} \min \left\{  | 2[\dot{\omega}_{\alpha_x} + x \dot{\alpha}_{x} + \alpha_{x} ] t - \dot{\beta}_{x} | : x \in \left(-1/\sqrt{2}, 1/\sqrt{2}\right) \right\} , \\
&& \liminf_{t \rightarrow \infty} \frac{L_t}{t} \geq  \frac{1}{\sqrt{2} \pi} \liminf_{t \rightarrow \infty}  \frac{ 
   \min \{ | 2[\dot{\omega}_{\alpha_x } + x \dot{\alpha}_{x} + \alpha_{x} ] t  - \dot{\beta}_{x} |: -1 \leq \sqrt{2}x \leq 1\} } {t} \\
&& \geq \frac{\sqrt{2}}{\pi} \min \left\{  |\dot{\omega}_{\alpha_x} + x \dot{\alpha}_{x} + \alpha_{x}   | : x \in \left(-1/\sqrt{2}, 1/\sqrt{2}\right) \right\}. 
\end{eqnarray*}

{\bf Alternative lower bound}: Instead of the lower bound in (\ref{n-Q*-P1-bound}), we may 
obtain an explicit  
lower bound for $c_0$ as follows: 
\[ c_0 \geq \frac{1}{2 \pi} \liminf_{t \rightarrow \infty} \int^{1/\sqrt{2}}_{-1/\sqrt{2}}  | H(x,t) | dx \geq \frac{1}{8 \sqrt{2} \pi} .\]
Indeed, using the first inequality of (\ref{n-H-bound}) we have 
\begin{eqnarray*} 
&& \int^{1/\sqrt{2}}_{-1/\sqrt{2}}  | H(x,t) | dx \geq \sum_{\ell=-L_t}^{L_t} \frac{|\xi_\ell | }{4} (x_{\ell+1} - x_{\ell}) \geq \frac{1}{4 \sqrt{2}} + O(t^{-1}), 
\end{eqnarray*}
where the last inequality is due to the following results: 
\begin{eqnarray*}
&& \xi_\ell - (x_\ell + x_{\ell+1})/2 = \xi_\ell - x_\ell - (x_{\ell+1} - x_\ell)/2 = O(t^{-1}), \\
&& \sum_{x_\ell \geq 0} \frac{|\xi_\ell | }{4} (x_{\ell+1} - x_{\ell}) = \frac{1}{8} \sum_{x_\ell \geq 0 } (x_\ell + x_{\ell+1})  (x_{\ell+1} - x_{\ell}) + O(t^{-1}) \\
&& = \frac{1}{8} \sum_{x_\ell \geq 0}  (x^2_{\ell+1} - x^2_{\ell}) + O(t^{-1}) = \frac{1}{8 \sqrt{2}} + O(t^{-1}), \\
&& \sum_{x_{\ell+1} \leq 0} \frac{|\xi_\ell | }{4} (x_{\ell+1} - x_{\ell}) = \frac{1}{8} \sum_{x_{\ell+1} \leq 0 } -(x_\ell + x_{\ell+1})  (x_{\ell+1} - x_{\ell}) + O(t^{-1}) \\
&& = \frac{1}{8} \sum_{x_{\ell+1} \geq 0}  (x^2_\ell - x^2_{\ell+1}) + O(t^{-1}) = \frac{1}{8 \sqrt{2}} + O(t^{-1}) . \\
\end{eqnarray*}

Finally, we proves (\ref{n-X-qc}) by plugging (\ref{n-Q-P2-epsilon})-(\ref{n-Q-Q*}) and (\ref{n-Q*-P1-bound}) into (\ref{n-X-qc-epsilon}).

\subsection*{A6. Proofs of (\ref{X-qc-epsilon}) and (\ref{X-qc-epsilon-1})} 

The distribution $\mP_{1t\eta}$ of $X^q_t(\eta)$ is an average of the distribution $\mP_{1t}$ of $X^q_t$, so $\mP_{1t\eta}$ is a convolution of 
$\mP_{1t}$ and a discrete uniform distribution. Since the convolution is a Markov kernel, we have $\gimel(X^q_t, X^c_t(\epsilon)) \leq \gimel(X^q_t(\eta), X^c_t(\epsilon))$, 
which is bounded by $\mho(X^q_t(\eta), X^c_t(\epsilon))$. Therefore, the result (\ref{mho-1}) in Theorem 1 immediately leads to (\ref{X-qc-epsilon}). 

The definitions of $\gimel$ and $\mho$ shows 
\[ \mho (X^q_t, X^c_t(\epsilon) ) = \max \{ \gimel(X^c_t(\epsilon), X^q_t), \gimel(X^q_t, X^c_t(\epsilon)) \} ,  \]
and hence (\ref{n-X-qc}) and (\ref{X-qc-epsilon}) together establish (\ref{X-qc-epsilon-1}).

\subsection*{A7. Proof of  (\ref{X-qc-eta})} 
It is enough to establish that both $\gimel(\mP_{1t\eta}, \mP_{2t})$ and 
$\gimel(\mP_{2t}, \mP_{1t\eta})$ are of order $\eta^{1/2}$ as $t \rightarrow \infty$. 
By the definition of $\mP_{2t}$, we have 
\begin{eqnarray*}
&& \limsup_{t \rightarrow \infty} \| \mP^*_{1t\eta} - \mP_{2t} \|_{TV} \leq 
  \limsup_{t \rightarrow \infty}   \limsup_{\epsilon \rightarrow 0}   \| \mP^*_{1t\eta} - \mP_{2t\epsilon} \|_{TV}  \\
&& \leq \limsup_{\epsilon \rightarrow 0}  \limsup_{t \rightarrow \infty}   \|\mP^*_{1t\eta} - \mP_{2t\epsilon} \|_{TV} \\
&& = O(\eta^{1/2}), 
\end{eqnarray*}
where the last equality is due to (\ref{equ-P*1P2}). Similarly, we have 
\begin{eqnarray*}
&& \limsup_{t \rightarrow \infty} \| \mP_{1t\eta} - \mP^*_{2t} \|_{TV} \leq 
  \limsup_{t \rightarrow \infty}   \limsup_{\epsilon \rightarrow 0}   \| \mP_{1t\eta} - \mP^*_{2t\epsilon} \|_{TV}  \\
&& \leq \limsup_{\epsilon \rightarrow 0}  \limsup_{t \rightarrow \infty}   \|\mP_{1t\eta} - \mP^*_{2t\epsilon} \|_{TV} \\
&& = O(\eta^{1/2}), 
\end{eqnarray*}
where the last equality is due to (\ref{equ-P1P*2}). 

With the above two results, we employ the same arguments for deriving (\ref{gimel12}) and (\ref{gimel21}) from (\ref{equ-P*1P2}) and (\ref{equ-P1P*2})
and obtain that as $t \rightarrow \infty$, $\gimel(\mP_{1t\eta}, \mP_{2t})$ and $\gimel(\mP_{2t}, \mP_{1t\eta})$ are of order $\eta^{1/2}$. 

\subsection*{A8. Proof of (\ref{X-qc})} 
The proof of (\ref{X-qc-eta}) has shown 
\[ \limsup_{t \rightarrow \infty} \| \mP^*_{1t\eta} - \mP_{2t} \|_{TV} = O(\eta^{1/2}), \] 
and again $\mP^*_{1t\eta}$ as a convolution of $\mP_{1t}$ and a uniform distribution, we have  
\[ \limsup_{t \rightarrow \infty} \gimel( \mP_{1t}, \mP_{2t} ) = O(\eta^{1/2}) . \]
Since the left-hand side of the above equation is free of $\eta$,  we let $\eta \rightarrow 0$ to prove the first result of (\ref{X-qc}). 
For the second result, we have 
\[ \gimel(X^c_t, X^q_t) \geq \gimel(X^c_t(\epsilon), X^q_t) - \mho(X^c_t, X^c_t(\epsilon)) . \]
Hence, we prove the second result of (\ref{X-qc}) as follows: 
\begin{eqnarray*}
&& \liminf_{t \rightarrow \infty} \gimel(X^c_t, X^q_t) \geq \liminf_{t \rightarrow \infty} \gimel(X^c_t(\epsilon), X^q_t) - \limsup_{t \rightarrow \infty} \mho(X^c_t, X^c_t(\epsilon)). \\
&&\geq c_0 - O(\epsilon^{1/2})   \rightarrow c_0 \mbox{ as } \epsilon \rightarrow 0 , 
\end{eqnarray*}
where we note that  the left-hand side is free of $\epsilon$, and 
the second inequality is due to (\ref{X-qc-epsilon-1}) and 
\[ 
\limsup_{t \rightarrow \infty} \mho(X^c_t, X^c_t(\epsilon)) = O(\epsilon^{1/2}) . \]
To show the above result, it is enough to prove 
\[ \limsup_{t \rightarrow \infty} \| \mP_{2t} - \mP_{2t\epsilon} \|_{TV} = O(\epsilon^{1/2}) . \]
In fact, we have 
\begin{eqnarray*}
&& \limsup_{t \rightarrow \infty} \| \mP_{2t} - \mP_{2t\epsilon} \|_{TV} \leq 
  \limsup_{t \rightarrow \infty}   \limsup_{\epsilon^\prime \rightarrow 0}   \| \mP_{2t\epsilon^\prime} - \mP_{2t\epsilon}  \|_{TV}  \\
&& \leq \limsup_{\epsilon^\prime \rightarrow 0}  \limsup_{t \rightarrow \infty}   \|\mP_{2t\epsilon^\prime} - \mP_ {2t\epsilon} \|_{TV} \\
&& = O(\epsilon^{1/2}), 
\end{eqnarray*}
where the first inequality is due the he definition of $\mP_{2t}$, and the last equality uses the fact that (\ref{equ-P2Q}) indicates 
\[  \limsup_{t \rightarrow \infty}   \|\mP_{2t\epsilon^\prime} - \mP_ {2t\epsilon} \|_{TV}  = O(\epsilon^{1/2} + (\epsilon^\prime) ^{1/2}) . \]

\end{document}